\documentstyle[amstex,amssymb]{article}

\catcode`\@=11
\def\numberbysection{\@addtoreset{equation}{section}
        \def\theequation{\thesection.\arabic{equation}}}
\numberbysection
\input{tcilatex}

\begin{document}

\newlength{\lno} \lno1.5cm \newlength{\len} \len=\textwidth%
\addtolength{\len}{-\lno}

\setcounter{page}{0}

\baselineskip7mm \renewcommand{\thefootnote}{\fnsymbol{footnote}} \newpage %
\setcounter{page}{0}

\begin{titlepage}     
\vspace{0.5cm}
\begin{center}
{\Large\bf Algebraic Bethe Ansatz for the Zamolodchikov-Fateev and Izergin-Korepin models with open boundary conditions}\\
\vspace{1cm}
{\large  V. Kurak $^{\dag }$\hspace{.5cm} and \hspace{.5cm} A. Lima-Santos$^{\ddag}$ } \\
\vspace{1cm}
$^{\dag}${\large \em Universidade de S\~ao Paulo, Instituto de F\'{\i}sica \\
Caixa Postal 66318, CEP 05315-970~~S\~ao Paulo -SP, Brasil}\\
\vspace{.5cm}
$^{\ddag}${\large \em Universidade Federal de S\~ao Carlos, Departamento de F\'{\i}sica \\
Caixa Postal 676, CEP 13569-905~~S\~ao Carlos, Brasil}\\
\end{center}
\vspace{1.2cm}

\begin{abstract}
We have considered the Zamolodchikov-Fateev and the Izergin-Korepin models with diagonal reflection boundaries. 
In each case the eigenspectrum of the transfer matrix is determined by application of the algebraic Bethe Ansatz .
\end{abstract}
\vspace{2cm}
\begin{center}
\end{center}
\vfill
\begin{center}
\small{\today}
\end{center}
\end{titlepage}

\baselineskip6mm

\newpage

\section{Introduction}

Some one-dimensional quantum spin chain Hamiltonians and some models of
classical statistical mechanics in two spatial dimensions on a lattice - the
vertex models - share a common mathematical structure responsible by
well-known breakthroughs in our present understanding of these models \cite%
{Baxter, KIB, ABR}. If the Boltzmann weights underlying the vertex models
are obtained from solutions of the Yang-Baxter ({\small YB}) equation, the
commutativity of a collection of transfer matrices depending on a spectral
parameter, say $u$, follows, leading to their integrability. Moreover, by
taking the logarithmic derivative of the transfer matrix, evaluated at a
special value of this spectral parameter, one gets an associated
one-dimensional quantum spin chain Hamiltonian. If fact, it might be said
that the most successful approach to construct integrable two-dimensional
lattice models of statistical mechanics is by solving the Yang-Baxter
equation. For a given solution of this equation, one can define local
Boltzmann weights to find a commuting family of transfer matrices.

The diagonalization of \ one-dimensional quantum spin chain Hamiltonians
started with the Bethe Ansatz ({\small BA}) \cite{Bethe}, which gradually
become a powerful method in the analysis of integrable models. There are
several versions: the coordinate {\small BA} \cite{Bethe}, the algebraic%
{\small \ BA} \cite{FT}, the analytical {\small BA} \cite{VR}, etc. The
simplest version is the coordinate {\small BA.}\ In this framework one can
obtain the eigenfunctions and the spectrum of the Hamiltonian from its
eigenvalue problem. It is really simple and clear for two-state models like
the six-vertex model but becomes tricky for models with a higher number of
states.

The development of the quantum inverse scattering method approach to
integrability resulted in the algebraic {\small BA} \cite{FT} which is an
elegant and important generalization of the coordinate {\small BA}. It is
based on the idea of constructing the eigenvectors of the transfer matrix
via the action of \ "creation " operators on a reference state. The creation
operators are just entries of the monodromy matrix whose trace is the
transfer matrix. Then by using the {\small YB} equation one can write the
so-called fundamental relation which entails the generalized \ "commutation
relations " of the transfer matrix with the remaining entries of the
monodromy matrix ( the \ "creation " and \ "annihilation " operators). The
creation operators that form a $n$-body Bethe vector are evaluated at some
unknown spots, say $\{u_{i}\}=\{u_{1},u_{2},...,u_{n}\}$. Finally, the
action of the transfer matrix, evaluated at some spectral parameter, say $u$%
, will be an eingenvalue problem, if a set of equations, known as Bethe
equations, not depending on $u$, determine the set $\{u_{i}\}$. Despite its
elegance and completeness, the actual implementation of the algebraic Bethe
Ansatz can become rather tricky and laborious, as will be exemplified in
this article. It is usual to rely on other methods such as the coordinate or
the analytical {\small BA } to gather important informations about the
eingenvalues of the transfer matrix.

The integrability of open spin chains in the framework of the quantum
inverse scattering method was formulated by Sklyanin relying on previous
results of Cherednik \cite{Che}. In reference \cite{Skl}, Sklyanin used his
formalism to solve the open spin-$1/2$ chain with diagonal boundary terms,
and his original formalism was successfully extended to others models by
Mezincescu and Nepomechie in \cite{MN}. The basic idea is similar to that of
the usual algebraic {\small BA}, with the important difference that now one
considers the so-called double-row monodromy matrix. When considering
systems on a finite interval, with independent boundary conditions at each
end, one has to introduce reflection matrices to describe such boundary
conditions. Integrable models with boundaries can be constructed out of a
pair of reflection $K$-matrices $K^{\pm }(u)$ which obey the cousin equation
of the {\small YB} equation - called reflection equation. $K^{-}(u)$ and $%
K^{+}(u)$ describe the effects of the presence of boundaries at the left and
the right ends, respectively. Sklyanin has shown that the double-row
monodromy matrix obeys a generalization of the fundamental relation also
named reflection equation. The non-diagonal entries of the double-row
monodromy matrix play the role of \ "creation " and "annihilation "
operators and the diagonal ones are used to construct the transfer matrix,
the reflection equation providing the fundamental set of \ "commutation
relations ". Similarly to the usual {\small BA} the creation operators
acting \ on a reference state form a $n$-body Bethe vector when evaluated at
some unknown spots, say $\{u_{i}\}=\{u_{1},u_{2},...,u_{n}\}$, and an
eingenvalue problem for the transfer will exist if a set of equations, named
also Bethe equations, not depending on $u$, determine the set $\{u_{i}\}$.%
\newline
\qquad In a recent article \cite{GLL} Guang-Liang Li, {\it et al }%
investigated the 19-vertex Izergin-Korepin ({\small IK}) model \cite{IK}
with boundaries using the algebraic {\small BA}. They have found a spurious
dependence of the Bethe equations on the spectral parameter $u$ , such that
the set $\{u_{i}\}$ could inconsistently depend on the spectral parameter$\
u $. Even more, this spurious dependence led the authors of \cite{GLL} to
raise the suspicion of the non-uniqueness (or inconsistency) of Skyanin's
algebraic {\small BA} for this model. On the other hand, this model has been
previously studied through the analytical {\small BA }\cite{MN3, YB}{\small %
\ }and the coordinate {\small BA} \cite{FLU} where no such spurious
dependence was found. So, we decided to investigate a related $19$-vertex
model , the Zamolodchikov-Fateev ({\small ZF}) model \cite{ZF}, where one
surely does not expect such a spurious dependence on $u$. The approach is
the same as that of \cite{GLL}, first envisaged by Fan \cite{FAN}, which is
the boundary version of the Tarasov work \cite{TA} for the periodic {\small %
IK} model. As a further check we also performed the calculations for the 
{\small IK} model .\newline
\qquad We do not have found such a spurious dependence for the {\small ZF}
model, our results agreeing with all previous results. For our pleasant
surprise we do not have found such a spurious dependence for the {\small IK}
model too. Indeed, our calculations agree with those of Guang-Liang Li, et
al and there is no spurious dependence on the spectral parameter at all. The
point is that this spurious dependence of the Bethe equations on the
spectral parameter in \cite{GLL} is only apparent. A careful analysis as we
perform em Section $4$ bellow shows that it does not exist. It can be also
checked that, in the notation of ref.\cite{GLL}, that their factor function $%
\beta (u,u_{i})$ in their Eq. ($97$) does not depend on $u$, so that the
dependence on u in their Eq ($102$), for the non-quantum group invariant
case, is non existent. The Sklyanin algebraic {\small BA} is unique and
consistent for the {\small ZF} and the {\small IK} vertex models.

The paper is organized as follows: in Section $2$ we define the models to be
studied and introduce the necessary notation to briefly review the basic
concepts of the algebraic {\small BA}.In section $3$, we present our
detailed calculations common to both models and in Section $4$ the
eigenspectra and the corresponding Bethe equations are explicitly presented
for each model. Section $5$ is reserved for conclusions and in the appendix
we derive the fundamental commutation relations for the double-row monodromy
matrix entries.

\section{The Models}

To determine an integrable vertex model on a lattice it is first necessary
that the bulk vertex weights be specified by a ${\cal R}$-matrix ${\cal R}%
(u) $, where $u$ is the spectral parameter. It acts on the tensor product $%
V\otimes V$ for a given vector space $V$ and satisfy a special system of
functional equations, the Yang-Baxter equation 
\begin{equation}
{\cal R}_{12}(u){\cal R}_{13}(u+v){\cal R}_{23}(v)={\cal R}_{23}(v){\cal R}%
_{13}(u+v){\cal R}_{12}(u),  \label{des.1}
\end{equation}%
in $V\otimes V\otimes V$, where ${\cal R}_{12}={\cal R}\otimes {\bf 1}$, $%
{\cal R}_{23}={\bf 1}\otimes {\cal R}$, etc.

The ${\cal R}$ matrix is said to be regular if it satisfies the property $%
{\cal R}(0)=P$, where $P$ is the permutation matrix in $V\otimes V$: $%
P(\left\vert \alpha \right\rangle \otimes \left\vert \beta \right\rangle
)=\left\vert \beta \right\rangle \otimes \left\vert \alpha \right\rangle $
for $\left\vert \alpha \right\rangle ,\left\vert \beta \right\rangle \in V$.
\ In addition, we will require \cite{MN} that ${\cal R}(u)$ satisfies the
following properties%
\begin{eqnarray}
{\rm regularity} &:&{\cal R}_{12}(0)=f(0)^{1/2}P_{12},  \nonumber \\
{\rm unitarity} &:&{\cal R}_{12}(u){\cal R}_{12}^{t_{1}t_{2}}(-u)=f(u), 
\nonumber \\
{\rm PT-symmetry} &:&P_{12}{\cal R}_{12}(u)P_{12}={\cal R}%
_{12}^{t_{1}t_{2}}(u),  \nonumber \\
{\rm cros}\text{sin}{\rm g-symmetry} &:&{\cal R}_{12}(u)=U_{1}{\cal R}%
_{12}^{t_{2}}(-u-\rho )U_{1}^{-1},  \label{des.2}
\end{eqnarray}%
where $f(u)=x_{1}(u)x_{1}(-u)$, $x_{1}(u)$ being defined for each model
bellow. $t_{i}$ denotes transposition in the space $i$ , $\rho $ is the
crossing parameter and $U$ determines the crossing matrix%
\begin{equation}
M=U^{t}U=M^{t}.  \label{des.3}
\end{equation}%
Unitarity and crossing-symmetry together imply the useful relation%
\begin{equation}
M_{1}{\cal R}_{12}^{t_{2}}(-u-\rho )M_{1}^{-1}{\cal R}_{12}^{t_{1}}(u-\rho
)=f(u).  \label{des.4}
\end{equation}

The boundary weights then follow from $K$-matrices which satisfy the
boundary versions of the Yang-Baxter equation \cite{Skl, MN}, the reflection
equation%
\begin{equation}
{\cal R}_{12}(u-v)K_{1}^{-}(u){\cal R}%
_{12}^{t_{1}t_{2}}(u+v)K_{2}^{-}(v)=K_{2}^{-}(v){\cal R}%
_{12}(u+v)K_{1}^{-}(u){\cal R}_{12}^{t_{1}t_{2}}(u-v),  \label{des.5}
\end{equation}%
and the dual reflection equation%
\[
{\cal R}_{12}(-u+v)(K_{1}^{+})^{t_{1}}(u)M_{1}^{-1}{\cal R}%
_{12}^{t_{1}t_{2}}(-u-v-2\rho )M_{1}(K_{2}^{+})^{t_{2}}(v) 
\]%
\begin{equation}
=(K_{2}^{+})^{t_{2}}(v)M_{1}{\cal R}_{12}(-u-v-2\rho
)M_{1}^{-1}(K_{1}^{+})^{t_{1}}(u){\cal R}_{12}^{t_{1}t_{2}}(-u+v).
\label{des.6}
\end{equation}%
In this case there is an isomorphism \ between $K^{-}$ and $K^{+}$ :%
\begin{equation}
K^{-}(u):\rightarrow K^{+}(u)=K^{-}(-u-\rho )^{t}M.  \label{des.7}
\end{equation}%
Therefore, given a solution of the reflection equation (\ref{des.5}) one can
also find a solution of the dual reflection equation (\ref{des.6}).

A quantum-integrable system is characterized by the monodromy matrix $T(u)$
satisfying the fundamental relation%
\begin{equation}
R(u-v)\left[ T(u)\otimes T(v)\right] =\left[ T(v)\otimes T(u)\right] R(u-v)
\label{des.7a}
\end{equation}%
where $R(u)$ is given by $R(u)=P{\cal R}(u).$

In the framework of the quantum inverse scattering method, the simplest
monodromies have become known as ${\cal L}$ operators, the Lax operators,
here defined by ${\cal L}_{aq}(u)={\cal R}_{aq}(u)$, where the subscript $a$
represents the auxiliary space, and $q$ represents the quantum space. The
monodromy matrix $T(u)$ is defined as the matrix product of $N$ \ Lax
operators on all sites of the lattice,%
\begin{equation}
T(u)={\cal L}_{aN}(u){\cal L}_{aN-1}(u)\cdots {\cal L}_{a1}(u).
\label{des.8}
\end{equation}

We recall that the main result for integrability is that, if the boundary
equations are satisfied, then the Sklyanin's transfer matrix%
\begin{equation}
t(u)={\rm Tr}_{a}\left( K^{+}(u)T(u)K^{-}(u)T^{-1}(-u)\right) ,
\label{des.9}
\end{equation}%
forms a commuting collection of operators in the quantum space%
\begin{equation}
\left[ t(u),t(v)\right] =0,\qquad \forall u,v  \label{des.10}
\end{equation}

The commutativity of $t(u)$ can be proved by using the unitarity and
crossing-unitarity relations, the reflection equation and the dual
reflection equation. In particular, it implies the integrability of an open
quantum spin chain whose Hamiltonian (with $K^{-}(0)=1$) is given by%
\begin{equation}
H=\sum_{k=1}^{N-1}H_{k,k+1}+\frac{1}{2}\left. \frac{dK_{1}^{-}(u)}{du}%
\right\vert _{u=0}+\frac{{\rm tr}_{0}K_{0}^{+}(0)H_{N,0}}{{\rm tr}K^{+}(0)},
\label{des.11}
\end{equation}%
where the two-site terms are given by%
\begin{equation}
H_{k,k+1}=\left. \frac{d}{du}P_{k,k+1}{\cal R}_{k,k+1}(u)\right\vert _{u=0},
\label{des.12}
\end{equation}%
in the standard fashion.

The three-state vertex models that we will consider are the
Zamolodchikov-Fateev model and the Izergin-Korepin model. Their ${\cal R}$%
-matrices have a common form%
\begin{equation}
{\cal R}(u)=\left( 
\begin{array}{ccc|ccc|ccc}
x_{1} &  &  &  &  &  &  &  &  \\ 
& x_{2} &  & x_{5} &  &  &  &  &  \\ 
&  & x_{3} &  & x_{6} &  & x_{7} &  &  \\ \hline
& y_{5} &  & x_{2} &  &  &  &  &  \\ 
&  & y_{6} &  & x_{4} &  & x_{6} &  &  \\ 
&  &  &  &  & x_{2} &  & x_{5} &  \\ \hline
&  &  &  &  &  &  &  &  \\[-10pt] 
&  & y_{7} &  & y_{6} &  & x_{3} &  &  \\ 
&  &  &  &  & y_{5} &  & x_{2} &  \\ 
&  &  &  &  &  &  &  & x_{1}%
\end{array}%
\right) ,  \label{des.18}
\end{equation}%
satisfying the properties (\ref{des.1}--\ref{des.4}).

In the context of the coordinate {\small BA}, these models were solved in 
\cite{FLU}, where an appropriate parametrization of wavefunctions was used \
in order to recast the coordinate {\small BA} for these three-states models
in a form as simple as the coordinate {\small BA} for the two-state models 
\cite{Alc}.

\subsection{\bf The Zamolodchikov-Fateev model}

This is the simplest three-state 19-vertex model \cite{ZF}. In the Bazhanov 
\cite{Baz} and Jimbo \cite{Jimbo} classification, it is the $B_{1}^{(1)}$
model or the $A_{1}^{(1)}$ model in the spin-$1$ representation \cite{KRS},
due to its construction from the six-vertex model by the fusion procedure.
The ${\cal R}$-matrix \ which satisfies the {\small YB} equation (\ref{des.1}%
) has the form (\ref{des.18}) with%
\begin{eqnarray}
x_{1}(u) &=&\sinh (u+\eta )\sinh (u+2\eta ),\quad x_{2}(u)=\sinh u\sinh
(u+\eta ),  \nonumber \\
x_{3}(u) &=&\sinh u\sinh (u-\eta ),\quad x_{4}(u)=\sinh u\sinh (u+\eta
)+\sinh \eta \sinh 2\eta ,  \nonumber \\
y_{5}(u) &=&x_{5}(u)=\sinh (u+\eta )\sinh 2\eta ,\quad
y_{6}(u)=x_{6}(u)=\sinh u\sinh 2\eta ,  \nonumber \\
y_{7}(u) &=&x_{7}(u)=\sinh \eta \sinh 2\eta .  \label{zf.1}
\end{eqnarray}%
This ${\cal R}$-matrix is regular and unitary, with $f(u)=x_{1}(u)x_{1}(-u)$%
, {\small P}- and {\small T}-symmetric and crossing-symmetric with $M=1$ and 
$\rho =\eta $. The most general diagonal solution for $K^{-}(u)$ has been
obtained in Ref. \cite{Mez} and is given by%
\begin{equation}
K^{-}(u,\beta )=\left( 
\begin{array}{ccc}
k_{11}^{-}(u) &  &  \\ 
& 1 &  \\ 
&  & k_{33}^{-}(u)%
\end{array}%
\right) ,  \label{zf.2}
\end{equation}%
with%
\begin{equation}
k_{11}^{-}(u)=-\frac{\beta \sinh u+2\cosh u}{\beta \sinh u-2\cosh u},\quad
k_{33}^{-}(u)=-\frac{\beta \sinh (u+\eta )-2\cosh (u+\eta )}{\beta \sinh
(u-\eta )+2\cosh (u-\eta )},  \label{zf.3}
\end{equation}%
where $\beta $ \ is a free parameter. By the automorphism (\ref{des.7}) the
solution for $K^{+}(u)$ follows 
\begin{equation}
K^{+}(u,\alpha )=K^{-}(-u-\rho ,\alpha )=\left( 
\begin{array}{ccc}
k_{11}^{+}(u) &  &  \\ 
& 1 &  \\ 
&  & k_{33}^{+}(u)%
\end{array}%
\right) ,  \label{zf.4}
\end{equation}%
with%
\begin{equation}
k_{11}^{+}(u)=-\frac{\alpha \sinh (u+\eta )-2\cosh (u+\eta )}{\alpha \sinh
(u+\eta )+2\cosh (u+\eta )},\quad k_{33}^{+}(u)=-\frac{\alpha \sinh u+2\cosh
u}{\alpha \sinh (u+2\eta )-2\cosh (u+2\eta )},  \label{zf.5}
\end{equation}%
where $\alpha $ is another free parameter.

The energy spectrum of this model was already obtained\ by Mezincescu {\it %
at al.} \cite{Mez} through a generalization of the quantum inverse
scattering method developed by Sklyanin \cite{Skl}, the so-called fusion
procedure \cite{KRS} and by Fireman et al \cite{FLU} through a
generalization of the coordinate {\small BA}.

For a particular choice of boundary terms, the {\small ZF} spin chain has
the quantum group symmetry {\it i.e.}, if we choose $\beta =2\coth \xi _{-}$
\ and $\ \alpha =2\coth \xi _{+}$, with $\xi _{\mp }\rightarrow \infty $,
then the open spin chain Hamiltonian has $U_{q}(su(2))$-invariance \cite{Mez}%
.

The fusion procedure was also used by Yung and Batchelor \cite{YB} to solve
the {\small ZF} vertex-model with inhomogeneities. The coordinate and
algebraic {\small BA} with periodic boundary conditions were presented in 
\cite{LS}.

\subsection{\bf The Izergin-Korepin model}

The solution of the {\small YB} equation corresponding to $A_{2}^{(2)}$ in
the fundamental representation was found by Izergin and Korepin \cite{IK}.
The ${\cal R}$-matrix has the form (\ref{des.18}) with non-zero entries%
\begin{eqnarray}
x_{1}(u) &=&\sinh (u-5\eta )+\sinh \eta ,\quad x_{2}(u)=\sinh (u-3\eta
)+\sinh 3\eta ,  \nonumber \\
x_{3}(u) &=&\sinh (u-\eta )+\sinh \eta ,\quad x_{4}(u)=\sinh (u-3\eta
)-\sinh 5\eta +\sinh 3\eta +\sinh \eta  \nonumber \\
x_{5}(u) &=&-2{\rm e}^{-u/2}\sinh 2\eta \cosh (\frac{u}{2}-3\eta ),\quad
y_{5}(u)=-2{\rm e}^{u/2}\sinh 2\eta \cosh (\frac{u}{2}-3\eta )  \nonumber \\
x_{6}(u) &=&2{\rm e}^{-u/2+2\eta }\sinh 2\eta \sinh (\frac{u}{2}),\quad
y_{6}(u)=-2{\rm e}^{u/2-2\eta }\sinh 2\eta \sinh (\frac{u}{2})  \nonumber \\
x_{7}(u) &=&-2{\rm e}^{-u+2\eta }\sinh \eta \sinh 2\eta -{\rm e}^{-\eta
}\sinh 4\eta ,\quad  \nonumber \\
y_{7}(u) &=&2{\rm e}^{u-2\eta }\sinh \eta \sinh 2\eta -{\rm e}^{\eta }\sinh
4\eta .
\end{eqnarray}%
This ${\cal R}$-matrix is regular and unitary, with $f(u)=x_{1}(u)x_{1}(-u)$%
. It is {\small PT}-symmetric and crossing-symmetric, with $\rho =-6\eta
-i\pi $ and%
\begin{equation}
M=\left( 
\begin{array}{ccc}
{\rm e}^{2\eta } &  &  \\ 
& 1 &  \\ 
&  & {\rm e}^{-2\eta }%
\end{array}%
\right) .  \label{ik.2}
\end{equation}

Diagonal solutions for $K^{-}(u)$ have been obtained in \cite{MN2}. It turns
out that there are three solutions without free parameters, being $%
K^{-}(u)=1 $, $K^{-}(u)=F^{+}$ and $K^{-}(u)=F^{-}$, with%
\begin{equation}
F^{\pm }=\left( 
\begin{array}{ccc}
{\rm e}^{-u}f^{\pm }(u) &  &  \\ 
& g^{\pm }(u) &  \\ 
&  & {\rm e}^{u}f^{\pm }(u)%
\end{array}%
\right) ,  \label{ik.3}
\end{equation}%
where we have defined%
\begin{equation}
f^{\pm }(u)=\cosh (\frac{1}{2}u-3\eta )\pm i\sinh (\frac{1}{2}u),\qquad
g^{\pm }(u)=\cosh (\frac{1}{2}u+3\eta )\mp i\sinh (\frac{1}{2}u)
\label{ik.4}
\end{equation}%
By the automorphism (\ref{des.7}), three solutions $K^{+}(u)$ follow as $%
K^{+}(u)=M$, $K^{+}(u)=G^{+}$ and $K^{+}(u)=G^{-}$, with%
\begin{equation}
G^{\pm }=\left( 
\begin{array}{ccc}
{\rm e}^{u-4\eta }f^{\pm }(u) &  &  \\ 
& h^{\pm }(u) &  \\ 
&  & {\rm e}^{-u+4\eta }f^{\pm }(u)%
\end{array}%
\right) ,  \label{ik.5}
\end{equation}%
where we have defined%
\begin{equation}
h^{\pm }(u)=\cosh (\frac{1}{2}u-3\eta )\mp i\sinh (\frac{1}{2}u-6\eta )).
\label{ik.6}
\end{equation}%
Here we will only consider three types of boundary solutions, one for each
pair ($K^{-}(u),K^{+}(u)$) defined by the automorphism (\ref{des.7}): ($1,M$%
), ($F^{+},G^{+}$) and ($F^{-},G^{-}$).

The transfer matrix for the case $(1,M)$ , whose corresponding spin-chain is 
$U_{q}(su(2))$-invariant \cite{MN4} has been diagonalized by the analytical 
{\small BA} in Ref. \cite{MN3}. For the other cases, ($F^{+},G^{+}$) and ($%
F^{-},G^{-}$) ,the corresponding open chain Hamiltonians are not $%
U_{q}(su(2))$-invariant . Nevertheless, their transfer matrices were
diagonalized by Yung and Batchelor in.\cite{YB} through the analytical 
{\small BA} with inhomogeneities. The energy eigenspectra for these three
cases was also solved by Fireman et al \cite{FLU}\ via the coordinate 
{\small BA}.

Recently has been argued by Nepomechie \cite{Nepo} that the transfer
matrices corresponding to these solutions also have the $U_{q}(o(3))$
symmetry, but with a nonstandard coproduct.

It was in this model that Tarasov developed the algebraic {\small BA} for
the three-state $19$-vertex models with periodic boundary conditions \cite%
{TA}. The analytical {\small BA} with periodic boundary conditions was
presented in \cite{VR} and the corresponding coordinate {\small BA} was
presented in \cite{LS}.

\section{Boundary Algebraic Bethe Ansatz}

In the previous section we have presented a common structure for the {\small %
ZF} and {\small IK} models. In this section we will turn to the eigenvalue
problem \ for their double-row transfer matrix with integrable boundaries,
named boundary algebraic Bethe Ansatz.

\subsection{The reference state}

The monodromy matrix $T(u)$ (\ref{des.8}) \ and its reflection $T^{-1}(-u)$
\ can be written as $3$ by $3$ matrices

\begin{equation}
T(u)=\left( 
\begin{array}{ccc}
T_{11}(u) & T_{12}(u) & T_{13}(u) \\ 
T_{21}(u) & T_{22}(u) & T_{23}(u) \\ 
T_{31}(u) & T_{32}(u) & T_{33}(u)%
\end{array}%
\right) ,\quad T^{-1}(-u)=\left( 
\begin{array}{ccc}
T_{11}^{-1}(-u) & T_{12}^{-1}(-u) & T_{13}^{-1}(-u) \\ 
T_{21}^{-1}(-u) & T_{22}^{-1}(-u) & T_{23}^{-1}(-u) \\ 
T_{31}^{-1}(-u) & T_{32}^{-1}(-u) & T_{33}^{-1}(-u)%
\end{array}%
\right)  \label{baba.1}
\end{equation}%
where 
\begin{equation}
T_{ia}(u)=\sum_{k_{1},...,k_{N-1}=1}^{3}{\cal L}_{ik_{1}}^{(N)}(u,\eta
)\otimes {\cal L}_{k_{1}k_{2}}^{(N-1)}(u,\eta )\otimes \cdots \otimes {\cal L%
}_{k_{N-1}a}^{(1)}(u,\eta )  \label{baba.2}
\end{equation}%
where ${\cal L}_{ij}^{(n)}$ are $3\times 3$ matrices acting on the $n${\rm th%
} site of the lattice, defined by 
\[
{\cal L}_{11}^{(n)}=\left( 
\begin{array}{ccc}
x_{1} & 0 & 0 \\ 
0 & x_{2} & 0 \\ 
0 & 0 & x_{3}%
\end{array}%
\right) ,\quad {\cal L}_{12}^{(n)}=\left( 
\begin{array}{ccc}
0 & 0 & 0 \\ 
x_{5} & 0 & 0 \\ 
0 & x_{6} & 0%
\end{array}%
\right) ,\quad {\cal L}_{13}^{(n)}=\left( 
\begin{array}{ccc}
0 & 0 & 0 \\ 
0 & 0 & 0 \\ 
x_{7} & 0 & 0%
\end{array}%
\right) , 
\]%
\[
{\cal L}_{21}^{(n)}=\left( 
\begin{array}{ccc}
0 & y_{5} & 0 \\ 
0 & 0 & y_{6} \\ 
0 & 0 & 0%
\end{array}%
\right) ,\quad {\cal L}_{22}^{(n)}=\left( 
\begin{array}{ccc}
x_{2} & 0 & 0 \\ 
0 & x_{4} & 0 \\ 
0 & 0 & x_{2}%
\end{array}%
\right) ,\quad {\cal L}_{23}^{(n)}=\left( 
\begin{array}{ccc}
0 & 0 & 0 \\ 
x_{6} & 0 & 0 \\ 
0 & x_{5} & 0%
\end{array}%
\right) , 
\]%
\begin{equation}
{\cal L}_{31}^{(n)}=\left( 
\begin{array}{ccc}
0 & 0 & y_{7} \\ 
0 & 0 & 0 \\ 
0 & 0 & 0%
\end{array}%
\right) ,\quad {\cal L}_{32}^{(n)}=\left( 
\begin{array}{ccc}
0 & y_{6} & 0 \\ 
0 & 0 & y_{5} \\ 
0 & 0 & 0%
\end{array}%
\right) ,\quad {\cal L}_{33}^{(n)}=\left( 
\begin{array}{ccc}
x_{3} & 0 & 0 \\ 
0 & x_{2} & 0 \\ 
0 & 0 & x_{1}%
\end{array}%
\right) .  \label{baba.3}
\end{equation}%
Using the unitary relation in (\ref{des.2}) we can see that the reflected
monodromy matrix $T^{-1}(-u)$ has the following matrix elements%
\begin{equation}
T_{bj}^{-1}(-u)=\frac{1}{f(u)^{N}}\sum_{k_{1},...,k_{N-1}=1}^{3}{\cal L}%
_{bk_{1}}^{(1)}(-u,-\eta )\otimes {\cal L}_{k_{1}k_{2}}^{(2)}(-u,-\eta
)\otimes \cdots \otimes {\cal L}_{k_{N-1}j}^{(N)}(-u,-\eta ).  \label{baba.4}
\end{equation}

Now we introduce the reference state%
\begin{equation}
\left\vert 0\right\rangle =\left( 
\begin{array}{c}
1 \\ 
0 \\ 
0%
\end{array}%
\right) _{(1)}\otimes \left( 
\begin{array}{c}
1 \\ 
0 \\ 
0%
\end{array}%
\right) _{(2)}\otimes \cdots \otimes \left( 
\begin{array}{c}
1 \\ 
0 \\ 
0%
\end{array}%
\right) _{(N)}  \label{baba.5}
\end{equation}%
The actions of $T(u)$ and $T^{-1}(-u)$ on this state are%
\begin{equation}
T(u)\left\vert 0\right\rangle =f^{N}(u)T^{-1}(-u)\left\vert 0\right\rangle
=\left( 
\begin{array}{ccc}
x_{1}^{N}(u)\left\vert 0\right\rangle & \ast & \ast \ast \\ 
0 & x_{2}^{N}(u)\left\vert 0\right\rangle & \ast \ast \ast \\ 
0 & 0 & x_{3}^{N}(u)\left\vert 0\right\rangle%
\end{array}%
\right)  \label{baba.6}
\end{equation}%
which give us, in the usual Bethe Ansatz language, the creation and
annihilation operators for this reference state. Moreover, we are dealing
with reflection and in this case we have a double-row monodromy matrix
defined by%
\begin{equation}
U(u)=T(u)K^{-}(u)T^{-1}(-u)=\left( 
\begin{array}{ccc}
U_{11}(u) & U_{12}(u) & U_{13}(u) \\ 
U_{21}(u) & U_{22}(u) & U_{23}(u) \\ 
U_{31}(u) & U_{32}(u) & U_{33}(u)%
\end{array}%
\right)  \label{baba.7}
\end{equation}%
where $K^{(-)}(u)$ is a reflection matrix.

For the diagonal case $K^{(-)}(u)={\rm diag}%
(k_{11}^{-}(u),k_{22}^{-}(u),k_{33}^{-}(u))$, the matrix elements of $U$
have the form 
\begin{equation}
U_{ij}(u)=\sum_{a=1}^{3}T_{ia}(u)k_{aa}^{-}(u)T_{aj}^{-1}(-u),\qquad
i,j=1,2,3.  \label{baba.8}
\end{equation}%
It follows from (\ref{baba.8}) that we will need to know the commutation
relations of the operators $T(u)$ and $T^{-1}(-u)$ in order to get the
action of $U(u)$ on the reference state (\ref{baba.5}). Using (\ref{des.7a})
with $u=-v$ we get the matrix relation%
\begin{equation}
T_{2}^{-1}(-u)R_{12}(2u)T_{1}(u)=T_{1}(u)R_{12}(2u)T_{2}^{-1}(-u)
\label{baba.9}
\end{equation}%
Applying both sides of this relation on the reference state, we find the
following relations between its matrix elements%
\begin{equation}
T_{21}(u)T_{12}^{-1}(-u)\left\vert 0\right\rangle =f_{1}(u)\frac{%
x_{1}^{2N}(u)-x_{2}^{2N}(u)}{f^{N}(u)}\left\vert 0\right\rangle
\label{baba.10}
\end{equation}%
\begin{equation}
T_{31}(u)T_{13}^{-1}(-u)\left\vert 0\right\rangle =f_{2}(u)\frac{%
x_{1}^{2N}(u)}{f^{N}(u)}\left\vert 0\right\rangle -f_{3}(u)f_{1}(u)\frac{%
x_{2}^{2N}(u)}{f^{N}(u)}\left\vert 0\right\rangle -f_{4}(u)\frac{%
x_{3}^{2N}(u)}{f^{N}(u)}\left\vert 0\right\rangle  \label{baba.11}
\end{equation}%
\begin{equation}
T_{32}(u)T_{23}^{-1}(-u)\left\vert 0\right\rangle =f_{2}(u)\frac{%
x_{2}^{2N}(u)-x_{3}^{2N}(u)}{f^{N}(u)}\left\vert 0\right\rangle
\label{baba.12}
\end{equation}%
where%
\begin{eqnarray}
f_{1}(u) &=&\frac{y_{5}(2u)}{x_{1}(2u)},\qquad f_{2}(u)=\frac{y_{7}(2u)}{%
x_{1}(2u)},\qquad f_{3}(u)=\frac{x_{1}(2u)y_{5}(2u)-x_{5}(2u)y_{7}(2u)}{%
x_{1}(2u)x_{4}(2u)-x_{5}(2u)y_{5}(2u)},  \nonumber \\
f_{4}(u) &=&\frac{x_{4}(2u)y_{7}(2u)-y_{5}^{2}(2u)}{%
x_{1}(2u)x_{4}(2u)-x_{5}(2u)y_{5}(2u)}.  \label{baba.13}
\end{eqnarray}%
Using these relations we can get the action of all $U_{ij}(u)$ on the
reference state%
\begin{eqnarray}
U_{11}(u)\left\vert 0\right\rangle &=&k_{11}^{-}(u)\frac{x_{1}^{2N}(u)}{%
f^{N}(u)}\left\vert 0\right\rangle  \nonumber \\
&&  \nonumber \\
U_{22}(u)\left\vert 0\right\rangle &=&f_{1}(u)U_{11}(u)\left\vert
0\right\rangle +\left[ k_{22}^{-}(u)-k_{11}^{-}(u)f_{1}(u)\right] \frac{%
x_{2}^{2N}(u)}{f^{N}(u)}\left\vert 0\right\rangle  \nonumber \\
&&  \nonumber \\
U_{33}(u)\left\vert 0\right\rangle &=&\left[ \left(
f_{2}(u)-f_{1}(u)f_{3}(u)\right) U_{11}(u)+f_{3}(u)U_{22}(u)\right]
\left\vert 0\right\rangle  \nonumber \\
&&+\left[ k_{33}^{-}(u)-k_{22}^{-}(u)f_{3}(u)-k_{11}^{-}(u)f_{4}(u)\right] 
\frac{x_{3}^{2N}(u)}{f^{N}(u)}\left\vert 0\right\rangle  \label{baba.14}
\end{eqnarray}%
and%
\begin{equation}
U_{ij}(u)\left\vert 0\right\rangle =0,\quad (i>j),\qquad U_{ij}(u)\left\vert
0\right\rangle \neq \left\{ 0,\left\vert 0\right\rangle \right\} ,\quad (i<j)
\label{baba.15}
\end{equation}%
In order to recover the usual {\small BA} structure we define news operators:%
\[
{\cal D}_{1}(u)=U_{11}(u),\qquad {\cal B}_{1}(u)=U_{12}(u),\qquad {\cal B}%
_{2}(u)=U_{13}(u) 
\]%
\[
{\cal C}_{1}(u)=U_{21}(u),\qquad {\cal D}_{2}(u)=U_{22}(u)-f_{1}(u){\cal D}%
_{1}(u),\qquad {\cal B}_{3}(u)=U_{23}(u) 
\]%
\begin{equation}
{\cal C}_{2}(u)=U_{31}(u),\qquad {\cal C}_{3}(u)=U_{32}(u),\qquad {\cal D}%
_{3}(u)=U_{33}(u)-f_{2}(u){\cal D}_{1}(u)-f_{3}(u){\cal D}_{2}(u)
\label{baba.16}
\end{equation}%
and then the new double-row monodromy matrix has the form%
\begin{equation}
U(u)\rightarrow {\cal U}(u)=\left( 
\begin{array}{ccc}
{\cal D}_{1}(u) & {\cal B}_{1}(u) & {\cal B}_{2}(u) \\ 
{\cal C}_{1}(u) & {\cal D}_{2}(u) & {\cal B}_{3}(u) \\ 
{\cal C}_{2}(u) & {\cal C}_{3}(u) & {\cal D}_{3}(u)%
\end{array}%
\right)  \label{baba.17}
\end{equation}%
The action of $\ {\cal U}(u)$ on the reference state is now 
\begin{equation}
{\cal U}(u)\left\vert 0\right\rangle =\left( 
\begin{array}{ccc}
{\cal X}_{1}(u)\left\vert 0\right\rangle & \ast & \ast \ast \\ 
0 & {\cal X}_{2}(u)\left\vert 0\right\rangle & \ast \ast \ast \\ 
0 & 0 & {\cal X}_{3}(u)\left\vert 0\right\rangle%
\end{array}%
\right)  \label{baba.18}
\end{equation}%
where%
\begin{eqnarray}
{\cal X}_{1}(u) &=&k_{11}^{-}(u)\frac{x_{1}^{2N}(u)}{f^{N}(u)}  \nonumber \\
{\cal X}_{2}(u) &=&\left[ k_{22}^{-}(u)-k_{11}^{-}(u)f_{1}(u)\right] \frac{%
x_{2}^{2N}(u)}{f^{N}(u)}  \nonumber \\
{\cal X}_{3}(u) &=&\left[
k_{33}^{-}(u)-k_{22}^{-}(u)f_{3}(u)-k_{11}^{-}(u)f_{4}(u)\right] \frac{%
x_{3}^{2N}(u)}{f^{N}(u)}  \label{baba.19}
\end{eqnarray}

The transfer matrix $t(u)$ (\ref{des.9}), with diagonal left reflection \ $%
K^{(+)}={\rm diag}(k_{11}^{+},k_{22}^{+}.k_{33}^{+})$ has the form%
\begin{eqnarray}
t(u) &=&k_{11}^{+}(u)U_{11}(u)+k_{22}^{+}(u)U_{22}(u)+k_{33}^{+}(u)U_{33}(u)
\nonumber \\
&=&\Omega _{1}(u){\cal D}_{1}(u)+\Omega _{2}(u){\cal D}_{2}(u)+\Omega _{3}(u)%
{\cal D}_{3}(u)  \label{baba.20}
\end{eqnarray}%
where%
\begin{eqnarray}
\Omega _{1}(u) &=&k_{11}^{+}(u)+f_{1}(u)k_{22}^{+}(u)+f_{2}(u)k_{33}^{+}(u) 
\nonumber \\
\Omega _{2}(u) &=&k_{22}^{+}(u)+f_{3}(u)k_{33}^{+}(u)  \nonumber \\
\Omega _{3}(u) &=&k_{33}^{+}(u)  \label{baba.21}
\end{eqnarray}

It is now clear that for ${\cal U}(u)$ we have recovered the usual algebraic 
{\small BA} structure. Therefore we can look for states created by the
action of the operators ${\cal B}_{i}(u)$ on the reference $\Psi _{0}$ which
will be eigenstates of the transfer matrix (\ref{des.9}). To do this we
first recall the magnon number operator%
\begin{equation}
M=\sum_{k=1}^{N}M_{k},\qquad M_{k}={\rm diag}(0,1,2)  \label{baba.22}
\end{equation}%
This is the analogue of the operator $S_{T}^{z}$ used in the coordinate 
{\small BA} construction. The relation $M\Psi _{m}=m\Psi _{m}$ where $%
m=N-S_{T}^{z}$, allows us to build states $\Psi _{m}$ such that $t(u)\Psi
_{m}=\Lambda _{m}\Psi _{m}$. Therefore, we can start the diagonalization of $%
t(u)$ by considering all possible values of $m$ in a lattice with $N$ sites.

By the previous construction, $\Psi _{0}$ \ is our reference state $|0>$ ,
which is itself an eigenstate of $t(u)$ 
\begin{equation}
t(u)\Psi _{0}=\Lambda _{0}(u)\Psi _{0}  \label{baba.23}
\end{equation}%
with eigenvalue%
\begin{eqnarray}
\Lambda _{0}(u) &=&\left[
k_{11}^{+}(u)+f_{1}(u)k_{22}^{+}(u)+f_{2}(u)k_{33}^{+}(u)\right]
k_{11}^{-}(u)\frac{x_{1}^{2N}(u)}{f^{N}(u)}  \nonumber \\
&&+\left[ k_{22}^{+}(u)+f_{3}(u)k_{33}^{+}(u)\right] \left[
k_{22}^{-}(u)-k_{11}^{-}(u)f_{1}(u)\right] \frac{x_{2}^{2N}(u)}{f^{N}(u)} 
\nonumber \\
&&+k_{33}^{+}(u)\left[
k_{33}^{-}(u)-k_{22}^{-}(u)f_{3}(u)-k_{11}^{-}(u)f_{4}(u)\right] \frac{%
x_{3}^{2N}(u)}{f^{N}(u)}  \label{baba.24}
\end{eqnarray}%
This is the only state with $m=0$.

\subsection{The one-particle state}

For $m=1$ we seek a state of the form%
\begin{equation}
\Psi _{1}(u_{1})={\cal B}_{1}(u_{1})\left\vert 0\right\rangle .
\label{baba.25}
\end{equation}%
Then the action of the transfer matrix $t(u)$ on this state is 
\begin{equation}
t(u)\Psi _{1}(u_{1})=\Omega _{1}(u){\cal D}_{1}(u){\cal B}%
_{1}(u_{1})\left\vert 0\right\rangle +\Omega _{2}(u){\cal D}_{2}(u){\cal B}%
_{1}(u_{1})\left\vert 0\right\rangle +\Omega _{3}(u){\cal D}_{3}(u){\cal B}%
_{1}(u_{1})\left\vert 0\right\rangle .  \label{baba.26}
\end{equation}%
Since we know the action of the operators ${\cal D}_{i}(u)$ on the reference
state $\left\vert 0\right\rangle $, we need to arrange the operator products 
\begin{equation}
{\cal D}_{1}(u){\cal B}_{1}(u_{1}),\ \quad {\cal D}_{2}(u){\cal B}%
_{1}(u_{1})\quad {\rm and}\quad {\cal D}_{3}(u){\cal B}_{1}(u_{1})
\label{baba.27}
\end{equation}%
in a normal-ordered form \cite{TA} .

We anticipate that, in general, the operator-valued function $\Psi
_{n}(u_{1},\ldots ,u_{n})$ for a $n$-particle Bethe state will be composed
by a set of normal-ordered monomials. A monomial is said to be in normal
order if all elements ${\cal B}_{i}$ are on the left, and all elements $%
{\cal C}_{i}$ are on the right of the elements ${\cal D}_{i}.$

In order to get this normal ordering we recall that the double-row monodromy
matrix $\ {\cal U}(u)$ satisfies the fundamental reflection equation 
\begin{equation}
{\cal R}_{12}(u-v){\cal U}_{1}(u){\cal R}_{21}(u+v){\cal U}_{2}(v)={\cal U}%
_{2}(v){\cal R}_{12}(u+v){\cal U}_{1}(u){\cal R}_{21}(u-v),  \label{baba.28}
\end{equation}%
where ${\cal U}_{1}(u)={\cal U}(u)\otimes 1,\ {\cal U}_{2}(u)=1\otimes {\cal %
U}(u)$ and ${\cal R}_{21}(u)=P{\cal R}_{12}(u)P$. In the appendix we show
how this equation (indeed a set of $81$ equations for the three-state
models) can be used to recast the non-normal ordered operator products as
the above into a linear combination of normal-ordered ones , which might be
called fundamental set of generalized commutation relations, or shortly
commutations relations.

For the present case $t(u)\Psi _{1}(u_{1})$ can be computed with the aid of
the following commutation relations (see the appendix)%
\begin{eqnarray}
{\cal D}_{1}(u){\cal B}_{1}(u_{1}) &=&a_{11}(u,u_{1}){\cal B}_{1}(u_{1})%
{\cal D}_{1}(u)+a_{12}(u,u_{1}){\cal B}_{1}(u){\cal D}%
_{1}(u_{1})+a_{13}(u,u_{1}){\cal B}_{1}(u){\cal D}_{2}(u_{1})  \nonumber \\
&&+a_{14}(u,u_{1}){\cal B}_{2}(u){\cal C}_{1}(u_{1})+a_{15}(u,u_{1}){\cal B}%
_{2}(u){\cal C}_{3}(u_{1})+a_{16}(u,u_{1}){\cal B}_{2}(u_{1}){\cal C}_{1}(u)
\label{baba.29}
\end{eqnarray}%
\begin{eqnarray}
{\cal D}_{2}(u){\cal B}_{1}(u_{1}) &=&a_{21}(u,u_{1}){\cal B}_{1}(u_{1})%
{\cal D}_{2}(u)+a_{22}(u,u_{1}){\cal B}_{1}(u){\cal D}%
_{1}(u_{1})+a_{23}(u,u_{1}){\cal B}_{1}(u){\em D}_{2}(u_{1})  \nonumber \\
&&+a_{24}(u,u_{1}){\cal B}_{3}(u){\cal D}_{1}(u_{1})+a_{25}(u,u_{1}){\cal B}%
_{3}(u){\cal D}_{2}(u_{1})+a_{26}(u,u_{1}){\cal B}_{2}(u){\cal C}_{1}(u_{1})
\nonumber \\
&&+a_{27}(u,u_{1}){\cal B}_{2}(u){\cal C}_{3}(u_{1})+a_{28}(u,u_{1}){\cal B}%
_{2}(u_{1}){\cal C}_{1}(u)+a_{29}(u,u_{1}){\cal B}_{2}(u_{1}){\cal C}_{3}(u)
\label{baba.30}
\end{eqnarray}%
\begin{eqnarray}
{\cal D}_{3}(u){\cal B}_{1}(u_{1}) &=&a_{31}(u,u_{1}){\cal B}_{1}(u_{1})%
{\cal D}_{3}(u)+a_{32}(u,u_{1}){\cal B}_{1}(u){\cal D}%
_{1}(u_{1})+a_{33}(u,u_{1}){\cal B}_{1}(u){\cal D}_{2}(u_{1})  \nonumber \\
&&+a_{34}(u,u_{1}){\cal B}_{3}(u){\cal D}_{1}(u_{1})+a_{35}(u,u_{1}){\cal B}%
_{3}(u){\cal D}_{2}(u_{1})+a_{36}(u,u_{1}){\cal B}_{2}(u){\cal C}_{1}(u_{1})
\nonumber \\
&&+a_{37}(u,u_{1}){\cal B}_{2}(u){\cal C}_{3}(u_{1})+a_{38}(u,u_{1}){\cal B}%
_{2}(u_{1}){\cal C}_{1}(u)+a_{39}(u,u_{1}){\cal B}_{2}(u_{1}){\cal C}_{3}(u)
\label{baba.31}
\end{eqnarray}%
Introducing these relations in the eq (\ref{baba.26}) one gets%
\[
t(u)\Psi _{1}(u_{1})=\Omega _{1}(u){\cal D}_{1}(u){\cal B}%
_{1}(u_{1})\left\vert 0\right\rangle +\Omega _{2}(u){\cal D}_{2}(u){\cal B}%
_{1}(u_{1})\left\vert 0\right\rangle +\Omega _{3}(u){\cal D}_{3}(u){\cal B}%
_{1}(u_{1})\left\vert 0\right\rangle 
\]%
\begin{eqnarray}
&=&\left[ a_{11}(u,u_{1})\Omega _{1}(u){\cal X}_{1}(u)+a_{21}(u,u_{1})\Omega
_{2}(u){\cal X}_{2}(u)+a_{31}(u,u_{1})\Omega _{3}(u){\cal X}_{3}(u)\right]
\Psi _{1}(u_{1})  \nonumber \\
&&+[{\cal X}_{1}(u_{1})\sum_{j=1}^{3}\Omega _{j}(u)a_{j2}(u,u_{1})+{\cal X}%
_{2}(u_{1})\sum_{j=1}^{3}\Omega _{j}(u)a_{j3}(u,u_{1})]B_{1}(u)\left\vert
0\right\rangle  \nonumber \\
&&+[{\cal X}_{1}(u_{1})\sum_{j=2}^{3}\Omega _{j}(u)a_{j4}(u,u_{1})+{\cal X}%
_{2}(u_{1})\sum_{j=2}^{3}\Omega _{j}(u)a_{j5}(u,u_{1})]B_{3}(u)\left\vert
0\right\rangle  \label{baba.32}
\end{eqnarray}%
So $\Psi _{1}(u_{1})$ will be an eigenstate of $t(u)$ with eigenvalue%
\begin{equation}
\Lambda _{1}(u,u_{1})=\sum_{j=1}^{3}\Omega _{j}(u){\cal X}%
_{j}(u)a_{j1}(u,u_{1})  \label{baba.33}
\end{equation}%
provided the following equations are satisfied%
\begin{equation}
\frac{{\cal X}_{1}(u_{1})}{{\cal X}_{2}(u_{1})}=-\frac{\sum_{j=1}^{3}\Omega
_{j}(u)a_{j3}(u,u_{1})}{\sum_{j=1}^{3}\Omega _{j}(u)a_{j2}(u,u_{1})}=-\frac{%
\sum_{j=2}^{3}\Omega _{j}(u)a_{j5}(u,u_{1})}{\sum_{j=2}^{3}\Omega
_{j}(u)a_{j4}(u,u_{1})}\equiv \Theta (u_{1})  \label{baba.34}
\end{equation}%
These are the Bethe equations for the one-particle state and the
corresponding Bethe root $u_{1}$ does not depend on $u$.

\subsection{The two-particle state}

For $m=2$ there are two linearly independent states ${\cal B}_{1}(u_{1})%
{\cal B}_{1}(u_{2})\left\vert 0\right\rangle $ and ${\cal B}%
_{2}(u_{1})\left\vert 0\right\rangle $. Therefore we seek for eigenstates of 
$t(u)$ in the form%
\begin{equation}
\Psi _{2}(u_{1},u_{2})={\cal B}_{1}(u_{1}){\cal B}_{1}(u_{2})\left\vert
0\right\rangle +{\cal B}_{2}(u_{1})\Gamma (u_{1},u_{2})\left\vert
0\right\rangle  \label{baba.35}
\end{equation}%
where $\Gamma (u_{1},u_{2})$ is an operator-valued function. Next we will
use the condition that $\Psi _{2}(u_{1},u_{2})$ must be normal-ordered to
find $\Gamma (u_{1},u_{2})$.

The first term in the right hand side of \ the eq (\ref{baba.35}) has its
normal-orderd form given by the commutation relation:%
\begin{eqnarray}
{\cal B}_{1}(u_{1}){\cal B}_{1}(u_{2}) &=&\omega (u_{1},u_{2})\left[ {\cal B}%
_{1}(u_{2}){\cal B}_{1}(u_{1})+G_{d_{1}}(u_{2},u_{1}){\cal B}_{2}(u_{2})%
{\cal D}_{1}(u_{1})+G_{d_{2}}(u_{2},u_{1}){\cal B}_{2}(u_{2}){\cal D}%
_{2}(u_{1})\right]  \nonumber \\
&&-G_{d_{1}}(u_{1},u_{2}){\cal B}_{2}(u_{1}){\cal D}%
_{1}(u_{2})-G_{d_{2}}(u_{1},u_{2}){\cal B}_{2}(u_{1}){\cal D}_{2}(u_{2})
\label{baba.36}
\end{eqnarray}%
where%
\begin{eqnarray}
&&\omega (u_{1},u_{2})=\frac{%
x_{3}(u_{1}-u_{2})x_{4}(u_{1}-u_{2})-x_{6}(u_{1}-u_{2})y_{6}(u_{1}-u_{2})}{%
x_{1}(u_{1}-u_{2})x_{3}(u_{1}-u_{2})}  \nonumber \\
&&\omega (u_{2},u_{1})\omega (u_{1},u_{2})=1  \label{baba.37}
\end{eqnarray}%
\begin{equation}
G_{d_{1}}(u_{1},u_{2})=-\frac{x_{6}(u_{1}-u_{2})}{x_{3}(u_{1}-u_{2})}\frac{%
x_{2}(2u_{2})}{x_{1}(2u_{2})}  \label{baba.38}
\end{equation}%
\begin{equation}
G_{d_{2}}(u_{1},u_{2})=\frac{x_{6}(u_{1}+u_{2})}{x_{2}(u_{1}+u_{2})}
\label{baba.39}
\end{equation}%
Here we have used the following identities valid for both models,%
\begin{equation}
\frac{y_{6}(-u)}{x_{3}(-u)}=-\frac{x_{3}(u)x_{6}(u)-x_{7}(u)y_{6}(u)}{%
x_{3}(u)x_{4}(u)-x_{6}(u)y_{6}(u)}  \label{baba.40}
\end{equation}%
and%
\begin{equation}
\frac{x_{2}(2u)}{x_{1}(2u)}=\frac{%
y_{5}(u-v)x_{2}(u+v)+x_{2}(u-v)x_{5}(u+v)f_{1}(u)}{y_{5}(u-v)x_{1}(u+v)}
\label{baba.41}
\end{equation}%
Now we can see that (\ref{baba.35}) is normal ordered if it satisfies the
condition%
\begin{equation}
\Psi _{2}(u_{2},u_{1})=\omega (u_{2},u_{1})\Psi _{2}(u_{1},u_{2})
\label{baba.42}
\end{equation}%
This condition fixes $\Gamma (u_{1},u_{2})$ and, by construction, the unique
candidate for the eigenstate of $t(u)$ in the $m=2$ \ case has the form%
\begin{equation}
\Psi _{2}(u_{1},u_{2})={\cal B}_{1}(u_{1}){\cal B}_{1}(u_{2})\left\vert
0\right\rangle +G_{d_{1}}(u_{1},u_{2}){\cal B}_{2}(u_{1}){\cal D}%
_{1}(u_{2})\left\vert 0\right\rangle +G_{d_{2}}(u_{1},u_{2}){\cal B}%
_{2}(u_{1}){\cal D}_{2}(u_{2})\left\vert 0\right\rangle  \label{baba.43}
\end{equation}%
The action of $t(u)$ on this state reads 
\begin{eqnarray}
&&t(u)\Psi _{2}(u_{1},u_{2})=\Omega _{1}(u){\cal D}_{1}(u){\cal B}_{1}(u_{1})%
{\cal B}_{1}(u_{2})\left\vert 0\right\rangle +\Omega _{2}(u){\cal D}_{2}(u)%
{\cal B}_{1}(u_{1}){\cal B}_{1}(u_{2})\left\vert 0\right\rangle  \nonumber \\
&&+\Omega _{3}(u){\cal D}_{3}(u){\cal B}_{1}(u_{1}){\cal B}%
_{1}(u_{2})\left\vert 0\right\rangle +G_{d_{1}}(u_{1},u_{2})\Omega _{1}(u)%
{\cal D}_{1}(u){\cal B}_{2}(u_{1}){\cal D}_{1}(u_{2})\left\vert
0\right\rangle  \nonumber \\
&&+G_{d_{1}}(u_{1},u_{2})\Omega _{2}(u){\cal D}_{2}(u){\cal B}_{2}(u_{1})%
{\cal D}_{1}(u_{2})\left\vert 0\right\rangle +G_{d_{1}}(u_{1},u_{2})\Omega
_{3}(u){\cal D}_{3}(u){\cal B}_{2}(u_{1}){\cal D}_{1}(u_{2})\left\vert
0\right\rangle  \nonumber \\
&&+G_{d_{2}}(u_{1},u_{2})\Omega _{1}(u){\cal D}_{1}(u){\cal B}_{2}(u_{1})%
{\cal D}_{2}(u_{2})\left\vert 0\right\rangle +G_{d_{2}}(u_{1},u_{2})\Omega
_{2}(u){\cal D}_{2}(u){\cal B}_{2}(u_{1}){\cal D}_{1}(u_{2})\left\vert
0\right\rangle  \nonumber \\
&&+G_{d_{2}}(u_{1},u_{2})\Omega _{3}(u){\cal D}_{3}(u){\cal B}_{2}(u_{1})%
{\cal D}_{2}(u_{2})\left\vert 0\right\rangle  \label{baba.44}
\end{eqnarray}%
In order to recast the right hand side of the above expression to its normal
ordered form we will need the following commutation relations, in addition
to those presented in the $m=1$ case (see the appendix):%
\begin{eqnarray}
{\cal D}_{1}(u){\cal B}_{2}(v) &=&b_{11}(u,v){\cal B}_{2}(v){\cal D}%
_{1}(u)+b_{12}(u,v){\cal B}_{2}(u){\cal D}_{1}(v)+b_{13}(u,v){\cal B}_{2}(u)%
{\cal D}_{2}(v)  \nonumber \\
&&+b_{14}(u,v){\cal B}_{2}(u){\cal D}_{3}(v)+b_{15}(u,v){\cal B}_{1}(u){\cal %
B}_{1}(v)+b_{16}(u,v){\cal B}_{1}(u){\cal B}_{3}(v)  \label{baba.45}
\end{eqnarray}%
\begin{eqnarray}
{\cal D}_{2}(u){\cal B}_{2}(v) &=&b_{21}(u,v){\cal B}_{2}(v){\cal D}%
_{2}(u)+b_{22}(u,v){\cal B}_{2}(u){\cal D}_{1}(v)+b_{23}(u,v){\cal B}_{2}(u)%
{\cal D}_{2}(v)  \nonumber \\
&&+b_{24}(u,v){\cal B}_{2}(u){\cal D}_{3}(v)+b_{25}(u,v){\cal B}_{1}(u){\cal %
B}_{1}(v)+b_{26}(u,v){\cal B}_{1}(u){\cal B}_{3}(v)  \nonumber \\
&&+b_{27}(u,v){\cal B}_{3}(u){\cal B}_{1}(v)+b_{28}(u,v){\cal B}_{3}(u){\cal %
B}_{3}(v)  \label{baba.46}
\end{eqnarray}%
\begin{eqnarray}
{\cal D}_{3}(u){\cal B}_{2}(v) &=&b_{31}(u,v){\cal B}_{2}(v){\cal D}%
_{3}(u)+b_{32}(u,v){\cal B}_{2}(u){\cal D}_{1}(v)+b_{33}(u,v){\cal B}_{2}(u)%
{\cal D}_{2}(v)  \nonumber \\
&&+b_{34}(u,v){\cal B}_{2}(u){\cal D}_{3}(v)+b_{35}(u,v){\cal B}_{1}(u){\cal %
B}_{1}(v)+b_{36}(u,v){\cal B}_{1}(u){\cal B}_{3}(v)  \nonumber \\
&&+b_{37}(u,v){\cal B}_{3}(u){\cal B}_{1}(v)+b_{38}(u,v){\cal B}_{3}(u){\cal %
B}_{3}(v)  \label{baba.47}
\end{eqnarray}%
\begin{eqnarray}
{\cal C}_{1}(u){\cal B}_{1}(v) &=&c_{11}(u,v){\cal B}_{1}(v){\cal C}%
_{1}(u)+c_{12}(u,v){\cal B}_{1}(v){\cal C}_{3}(v)+c_{13}(u,v){\cal B}_{1}(u)%
{\cal C}_{3}(v)  \nonumber \\
&&+c_{14}(u,v){\cal B}_{2}(u){\cal C}_{3}(v)+c_{15}(u,v){\cal B}_{2}(v){\cal %
C}_{2}(u)+c_{16}(u,v){\cal D}_{1}(v){\cal D}_{1}(u)  \nonumber \\
&&+c_{17}(u,v){\cal D}_{1}(v){\cal D}_{2}(u)+c_{18}(u,v){\cal D}_{1}(u){\cal %
D}_{1}(v)+c_{19}(u,v){\cal D}_{1}(u){\cal D}_{2}(v)  \nonumber \\
&&+c_{110}(u,v){\cal D}_{2}(u){\cal D}_{1}(v)+c_{111}(u,v){\cal D}_{2}(u)%
{\cal D}_{2}(v)  \label{baba.48}
\end{eqnarray}%
\begin{eqnarray}
{\cal C}_{3}(u){\cal B}_{1}(v) &=&c_{21}(u,v){\cal B}_{1}(v){\cal C}%
_{1}(u)+c_{22}(u,v){\cal B}_{1}(v){\cal C}_{3}(v)+c_{23}(u,v){\cal B}_{1}(u)%
{\cal C}_{3}(v)  \nonumber \\
&&+c_{24}(u,v){\cal B}_{2}(u){\cal C}_{3}(v)+c_{25}(u,v){\cal B}_{2}(v){\cal %
C}_{2}(u)+c_{26}(u,v){\cal D}_{1}(v){\cal D}_{1}(u)  \nonumber \\
&&+c_{27}(u,v){\cal D}_{1}(v){\cal D}_{2}(u)+c_{28}(u,v){\cal D}_{1}(u){\cal %
D}_{3}(v)+c_{29}(u,v){\cal D}_{1}(u){\cal D}_{2}(v)  \nonumber \\
&&+c_{210}(u,v){\cal D}_{1}(u){\cal D}_{2}(v)+c_{211}(u,v){\cal D}_{2}(u)%
{\cal D}_{1}(v)+c_{212}(u,v){\cal D}_{2}(u){\cal D}_{2}(v)  \nonumber \\
&&+c_{213}(u,v){\cal D}_{3}(u){\cal D}_{1}(v)+c_{214}(u,v){\cal D}_{3}(u)%
{\cal D}_{1}(v))  \label{baba.49}
\end{eqnarray}%
After a straightforward calculation we obtain%
\[
t(u)\Psi _{2}(u_{1},u_{2})=[\sum_{j=1}^{3}\Omega _{j}(u){\cal X}%
_{j}(u)a_{j1}(u,u_{1})a_{j1}(u,u_{2})]\Psi _{2}(u_{1},u_{2}) 
\]%
\[
+[a_{11}(u_{1},u_{2}){\cal X}_{1}(u_{1})\sum_{j=1}^{3}\Omega
_{j}(u)a_{j2}(u,u_{1})+a_{21}(u_{1},u_{2}){\cal X}_{2}(u_{1})\sum_{j=1}^{3}%
\Omega _{j}(u)a_{j3}(u,u_{1})]{\cal B}_{1}(u){\cal B}_{1}(u_{2})\left\vert
0\right\rangle 
\]%
\[
+[a_{11}(u_{1},u_{2}){\cal X}_{1}(u_{1})\sum_{j=2}^{3}\Omega
_{j}(u)a_{j4}(u,u_{1})+a_{21}(u_{1},u_{2}){\cal X}_{2}(u_{1})\sum_{j=2}^{3}%
\Omega _{j}(u)a_{j5}(u,u_{1})]{\cal B}_{3}(u){\cal B}_{1}(u_{2})\left\vert
0\right\rangle 
\]%
\[
+[a_{11}(u_{2},u_{1}){\cal X}_{1}(u_{2})\sum_{j=1}^{3}\Omega
_{j}(u)a_{j2}(u,u_{2})+a_{21}(u_{2},u_{1}){\cal X}_{2}(u_{2})\sum_{j=1}^{3}%
\Omega _{j}(u)a_{j3}(u,u_{2})]\omega (u_{1},u_{2}){\cal B}_{1}(u){\em B}%
_{1}(u_{1})\left\vert 0\right\rangle 
\]%
\[
+[a_{11}(u_{2},u_{1}){\cal X}_{1}(u_{2})\sum_{j=2}^{3}\Omega
_{j}(u)a_{j4}(u,u_{2})+a_{21}(u_{2},u_{1}){\cal X}_{2}(u_{2})\sum_{j=2}^{3}%
\Omega _{j}(u)a_{j5}(u,u_{2})]\omega (u_{1},u_{2}){\cal B}_{3}(u){\cal B}%
_{1}(u_{1})\left\vert 0\right\rangle 
\]%
\begin{eqnarray}
&&+[{\cal X}_{1}(u_{1}){\cal X}_{1}(u_{2})\sum_{j=1}^{3}\Omega
_{j}(u)H_{j1}(u_{1},u_{2})+{\cal X}_{1}(u_{1}){\cal X}_{2}(u_{2})%
\sum_{j=1}^{3}\Omega _{j}(u)H_{j3}(u_{1},u_{2})  \nonumber \\
&&+{\cal X}_{2}(u_{1}){\cal X}_{1}(u_{2})\sum_{j=1}^{3}\Omega
_{j}(u)H_{j2}(u_{1},u_{2})+{\cal X}_{2}(u_{1}){\cal X}_{2}(u_{2})%
\sum_{j=1}^{3}\Omega _{j}(u)H_{j4}(u_{1},u_{2})]{\cal B}_{2}(u)\left\vert
0\right\rangle  \label{baba.50}
\end{eqnarray}%
where%
\begin{eqnarray}
H_{11}(u_{1},u_{2}) &=&a_{14}(u,u_{1})\left(
c_{16}(u_{1},u_{2})+c_{18}(u_{1},u_{2})\right) +a_{15}(u,u_{1})\left(
c_{26}(u_{1},u_{2})+c_{29}(u_{1},u_{2})\right)  \nonumber \\
&&+b_{12}(u,u_{1})G_{d_{1}}(u_{1},u_{2})+\omega
(u_{1},u)a_{11}(u,u_{1})a_{12}(u,u_{2})G_{d_{1}}(u,u_{1})  \nonumber \\
H_{12}(u_{1},u_{2}) &=&a_{14}(u,u_{1})\left(
c_{17}(u_{1},u_{2})+c_{110}(u_{1},u_{2})\right) +a_{15}(u,u_{1})\left(
c_{27}(u_{1},u_{2})+c_{211}(u_{1},u_{2})\right)  \nonumber \\
&&+b_{13}(u,u_{1})G_{d_{1}}(u_{1},u_{2})+\omega
(u_{1},u)a_{11}(u,u_{1})a_{12}(u,u_{2})G_{d_{2}}(u,u_{1})  \nonumber \\
H_{13}(u_{1},u_{2})
&=&a_{14}(u,u_{1})c_{19}(u_{1},u_{2})+a_{15}(u,u_{1})c_{210}(u_{1},u_{2})+b_{12}(u,u_{1})G_{d_{2}}(u_{1},u_{2})
\nonumber \\
&&+\omega (u_{1},u)a_{11}(u,u_{1})a_{13}(u,u_{2})G_{d_{1}}(u,u_{1}) 
\nonumber \\
H_{14}(u_{1},u_{2})
&=&a_{14}(u,u_{1})c_{111}(u_{1},u_{2})+a_{15}(u,u_{1})c_{212}(u_{1},u_{2})+b_{13}(u,u_{1})G_{d_{2}}(u_{1},u_{2})
\nonumber \\
&&+\omega (u_{1},u)a_{11}(u,u_{1})a_{13}(u,u_{2})G_{d_{2}}(u,u_{1})
\label{baba.51}
\end{eqnarray}%
and 
\begin{eqnarray}
H_{j1}(u_{1},u_{2}) &=&a_{j6}(u,u_{1})\left(
c_{16}(u_{1},u_{2})+c_{18}(u_{1},u_{2})\right) +a_{j7}(u,u_{1})\left(
c_{26}(u_{1},u_{2})+c_{29}(u_{1},u_{2})\right)  \nonumber \\
&&+b_{j2}(u,u_{1})G_{d_{1}}(u_{1},u_{2})+\omega
(u_{1},u)a_{j1}(u,u_{1})a_{j2}(u,u_{2})G_{d_{1}}(u,u_{1})  \nonumber \\
&&+a_{j1}(u,u_{1})a_{j4}(u,u_{2})d_{13}(u_{1},u)  \nonumber \\
H_{j2}(u_{1},u_{2}) &=&a_{j6}(u,u_{1})\left(
c_{17}(u_{1},u_{2})+c_{110}(u_{1},u_{2})\right) +a_{j7}(u,u_{1})\left(
c_{27}(u_{1},u_{2})+c_{211}(u_{1},u_{2})\right)  \nonumber \\
&&+b_{j3}(u,u_{1})G_{d_{1}}(u_{1},u_{2})+\omega
(u_{1},u)a_{j1}(u,u_{1})a_{j2}(u,u_{2})G_{d_{2}}(u,u_{1})  \nonumber \\
&&+a_{j1}(u,u_{1})a_{j4}(u,u_{2})d_{14}(u_{1},u)  \nonumber \\
H_{j3}(u_{1},u_{2})
&=&a_{j6}(u,u_{1})c_{19}(u_{1},u_{2})+a_{j7}(u,u_{1})c_{210}(u_{1},u_{2})+b_{j2}(u,u_{1})G_{d_{2}}(u_{1},u_{2})
\nonumber \\
&&+\omega
(u_{1},u)a_{j1}(u,u_{1})a_{j3}(u,u_{2})G_{d_{1}}(u,u_{1})+a_{j1}(u,u_{1})a_{j5}(u,u_{2})d_{13}(u_{1},u)
\nonumber \\
H_{j4}(u_{1},u_{2})
&=&a_{j6}(u,u_{1})c_{111}(u_{1},u_{2})+a_{j7}(u,u_{1})c_{212}(u_{1},u_{2})+b_{j3}(u,u_{1})G_{d_{2}}(u_{1},u_{2})
\nonumber \\
&&+\omega
(u_{1},u)a_{j1}(u,u_{1})a_{j3}(u,u_{2})G_{d_{2}}(u,u_{1})+a_{j1}(u,u_{1})a_{j5}(u,u_{2})d_{14}(u_{1},u)
\label{baba.52}
\end{eqnarray}%
for $j=2,3$.\newline
Again, $\Psi _{2}(u_{1},u_{2})$ will be an eigenstate of $t(u)$ with
eigenvalue%
\begin{equation}
\Lambda _{2}(u,u_{1},u_{2})=\sum_{j=1}^{3}\Omega _{j}(u){\cal X}%
_{j}(u)a_{j1}(u,u_{1})a_{j1}(u,u_{2})  \label{baba.53}
\end{equation}%
provided the following Bethe equations are satisfied%
\begin{equation}
\frac{{\cal X}_{1}(u_{1})}{{\cal X}_{2}(u_{1})}=\Theta (u_{1})\frac{%
a_{21}(u_{1},u_{2})}{a_{11}(u_{1},u_{2})},\qquad \frac{{\cal X}_{1}(u_{2})}{%
{\cal X}_{2}(u_{2})}=\Theta (u_{2})\frac{a_{21}(u_{2},u_{1})}{%
a_{11}(u_{2},u_{1})}\qquad  \label{baba.54}
\end{equation}%
where $\Theta (u_{i}),\ i=1,2$ are given by\ (\ref{baba.34}).

\subsection{The three-particle state}

In this case we have to consider combinations and permutations of the states
of the type ${\cal B}_{1}{\cal B}_{1}{\cal B}_{1}\left\vert 0\right\rangle $
and ${\cal B}_{2}{\cal B}_{1}\left\vert 0\right\rangle .$

Let us consider the normal-ordered operator 
\begin{equation}
\Phi _{3}(u_{1},u_{2},u_{3})={\cal B}_{1}(u_{1})\Phi _{2}(u_{2},u_{3})+{\cal %
B}_{2}(u_{1})\Phi _{1}(u_{2})\Gamma _{1}(u_{1},u_{2},u_{3})+{\cal B}%
_{2}(u_{1})\Phi _{1}(u_{3})\Gamma _{2}(u_{1},u_{2},u_{3})  \label{baba.55}
\end{equation}%
with two exchange proprieties%
\begin{equation}
\Phi _{3}(u_{1},u_{2},u_{3})=\omega (u_{1},u_{2})\Phi
_{3}(u_{2},u_{1},u_{3})=\omega (u_{2},u_{3})\Phi _{3}(u_{1},u_{3},u_{2}),
\label{baba.56}
\end{equation}%
which can be used, together the commutations relations, to find the operator
valued functions%
\begin{eqnarray}
\Gamma _{1}(u_{1},u_{2},u_{3}) &=&\omega (u_{2},u_{3})\left[
a_{11}(u_{3},u_{2})G_{d_{1}}(u_{1},u_{3}){\cal D}%
_{1}(u_{3})+a_{21}(u_{3},u_{2})G_{d_{2}}(u_{1},u_{3}){\cal D}_{2}(u_{3})%
\right]  \nonumber \\
\Gamma _{2}(u_{1},u_{2},u_{3}) &=&a_{11}(u_{2},u_{3})G_{d_{1}}(u_{1},u_{2})%
{\cal D}_{1}(u_{2})+a_{21}(u_{2},u_{3})G_{d_{2}}(u_{1},u_{2}){\cal D}%
_{2}(u_{2})  \label{baba.57}
\end{eqnarray}%
Note that%
\begin{equation}
\Gamma _{1}(u_{1},u_{2},u_{3})=\omega (u_{2},u_{3})\Gamma
_{2}(u_{1},u_{3},u_{2})  \label{baba.58}
\end{equation}

The eigenvalue problem for the $3$-particle state is now reduced to the
problem of finding the action of the operators ${\cal D}_{\alpha }(u)$ , $%
\alpha =1,2,3$ on the state%
\begin{equation}
\Psi _{3}(u_{1},u_{2},u_{3})=\Phi _{3}(u_{1},u_{2},u_{3})\left\vert
0\right\rangle  \label{baba.59}
\end{equation}%
To do this we will need recall the appendix to get more five commutation
relations: 
\begin{eqnarray}
{\cal D}_{1}(u){\cal B}_{3}(v) &=&x_{11}(u,v){\cal B}_{3}(v){\cal D}%
_{1}(u)+x_{12}(u,v){\cal B}_{1}(v){\cal D}_{1}(u)+x_{13}(u,v){\cal B}_{1}(u)%
{\cal D}_{1}(v)  \nonumber \\
&&+x_{14}(u,v){\cal B}_{1}(u){\cal D}_{2}(v)+x_{15}(u,v){\cal B}_{1}(u){\cal %
D}_{3}(v)+x_{16}(u,v){\cal B}_{2}(u){\cal C}_{1}(v)  \nonumber \\
&&+x_{17}(u,v){\cal B}_{2}(u){\cal C}_{3}(v)+x_{18}(u,v){\cal B}_{2}(v){\cal %
C}_{1}(u)  \label{baba.60}
\end{eqnarray}%
\begin{eqnarray}
{\cal D}_{2}(u){\cal B}_{3}(v) &=&x_{21}(u,v){\cal B}_{3}(v){\cal D}%
_{2}(u)+x_{22}(u,v){\cal B}_{1}(v){\cal D}_{2}(u)+x_{23}(u,v){\cal B}_{1}(u)%
{\cal D}_{1}(v)  \nonumber \\
&&+x_{24}(u,v){\cal B}_{1}(u){\cal D}_{2}(v)+x_{25}(u,v){\cal B}_{1}(u){\cal %
D}_{3}(v)+x_{26}(u,v){\cal B}_{3}(u){\cal D}_{1}(v)  \nonumber \\
&&+x_{27}(u,v){\cal B}_{3}(u){\cal D}_{2}(v)+x_{28}(u,v){\cal B}_{3}(u){\cal %
D}_{3}(v)+x_{29}(u,v){\cal B}_{2}(u){\cal C}_{1}(v)  \nonumber \\
&&+x_{210}(u,v){\cal B}_{2}(u){\cal C}_{3}(v)+x_{211}(u,v){\cal B}_{2}(v)%
{\cal C}_{1}(u)+x_{212}(u,v){\cal B}_{2}(v){\cal C}_{3}(u)  \label{baba.61}
\end{eqnarray}%
\begin{eqnarray}
{\cal D}_{3}(u){\cal B}_{3}(v) &=&x_{31}(u,v){\cal B}_{3}(v){\cal D}%
_{3}(u)+x_{32}(u,v){\cal B}_{1}(v){\cal D}_{3}(u)+x_{33}(u,v){\cal B}_{1}(u)%
{\cal D}_{1}(v)  \nonumber \\
&&+x_{34}(u,v){\cal B}_{1}(u){\cal D}_{2}(v)+x_{35}(u,v){\cal B}_{1}(u){\cal %
D}_{3}(v)+x_{36}(u,v){\cal B}_{3}(u){\cal D}_{1}(v)  \nonumber \\
&&+x_{37}(u,v){\cal B}_{3}(u){\cal D}_{2}(v)+x_{38}(u,v){\cal B}_{3}(u){\cal %
D}_{3}(v)+x_{39}(u,v){\cal B}_{2}(u){\cal C}_{1}(v)  \nonumber \\
&&+x_{310}(u,v){\cal B}_{2}(u){\cal C}_{3}(v)+x_{311}(u,v){\cal B}_{2}(v)%
{\cal C}_{1}(u)+x_{312}(u,v){\cal B}_{2}(v){\cal C}_{3}(u)  \label{baba.62}
\end{eqnarray}%
\begin{eqnarray}
{\cal C}_{1}(u){\cal B}_{2}(v) &=&y_{11}(u,v){\cal B}_{2}(v){\cal C}%
_{1}(u)+y_{12}(u,v){\cal B}_{2}(v){\cal C}_{3}(u)+y_{13}(u,v){\cal B}_{2}(u)%
{\cal C}_{1}(v)  \nonumber \\
&&+y_{14}(u,v){\cal B}_{2}(u){\cal C}_{3}(v)+y_{15}(u,v){\cal B}_{1}(v){\cal %
D}_{1}(u)+y_{16}(u,v){\cal B}_{1}(v){\cal D}_{2}(u)  \nonumber \\
&&+y_{17}(u,v){\cal B}_{3}(v){\cal D}_{1}(u)+y_{18}(u,v){\cal B}_{3}(v){\cal %
D}_{2}(u)+y_{19}(u,v){\cal B}_{1}(u){\cal D}_{1}(v)  \nonumber \\
&&+y_{110}(u,v){\cal B}_{1}(u){\cal D}_{2}(v)+y_{111}(u,v){\cal B}_{1}(u)%
{\cal D}_{3}(v)+y_{112}(u,v){\cal B}_{3}(u){\cal D}_{1}(v)  \nonumber \\
&&+y_{113}(u,v){\cal B}_{3}(u){\cal D}_{2}(v)+y_{114}(u,v){\cal B}_{3}(u)%
{\cal D}_{3}(v)  \label{baba.63}
\end{eqnarray}%
\begin{eqnarray}
{\cal C}_{3}(u){\cal B}_{2}(v) &=&y_{21}(u,v){\cal B}_{2}(v){\cal C}%
_{1}(u)+y_{22}(u,v){\cal B}_{2}(v){\cal C}_{3}(u)+y_{23}(u,v){\cal B}_{2}(u)%
{\cal C}_{1}(v)  \nonumber \\
&&+y_{24}(u,v){\cal B}_{2}(u){\cal C}_{3}(v)+y_{25}(u,v){\cal B}_{1}(v){\cal %
D}_{1}(u)+y_{26}(u,v){\cal B}_{1}(v){\cal D}_{2}(u)  \nonumber \\
&&+y_{27}(u,v){\cal B}_{1}(v){\cal D}_{3}(u)+y_{28}(u,v){\cal B}_{3}(v){\cal %
D}_{1}(u)+y_{29}(u,v){\cal B}_{3}(v){\cal D}_{2}(u)  \nonumber \\
&&+y_{210}(u,v){\cal B}_{3}(v){\cal D}_{3}(u)+y_{211}(u,v){\cal B}_{1}(u)%
{\cal D}_{1}(v)+y_{212}(u,v){\cal B}_{1}(u){\cal D}_{2}(v)  \nonumber \\
&&+y_{213}(u,v){\cal B}_{1}(u){\cal D}_{3}(v)+y_{214}(u,v){\cal B}_{3}(u)%
{\cal D}_{1}(v)+y_{215}(u,v){\cal B}_{3}(u){\cal D}_{2}(v)  \nonumber \\
&&+y_{216}(u,v){\cal B}_{3}(u){\cal D}_{3}(v)  \label{baba.64}
\end{eqnarray}%
After a straightforward but (the reader should be advised) quite lengthy
computation we obtain the following simplified expressions

\[
{\cal D}_{\alpha }(u)\Psi _{3}(u_{1},u_{2},u_{3})={\cal X}_{\alpha
}(u)\dprod\limits_{i=1}^{3}a_{\alpha 1}(u,u_{i})\Psi _{3}(u_{1},u_{2},u_{3}) 
\]%
\[
+\sum_{i=1}^{3}\dprod\limits_{j=1}^{i-1}\omega (u_{j},u_{i})[{\cal X}%
_{1}(u_{i})a_{\alpha 2}(u,u_{i})\dprod\limits_{k\neq
i}^{3}a_{11}(u_{i},u_{k})+{\cal X}_{2}(u_{i})a_{\alpha
3}(u,u_{i})\dprod\limits_{k\neq i}^{3}a_{21}(u_{i},u_{k})]{\cal B}%
_{1}(u)\Psi _{2}(\overset{\vee }{u_{i}}) 
\]%
\[
+(1-\delta _{\alpha ,1})\sum_{i=1}^{3}\dprod\limits_{j=1}^{i-1}\omega
(u_{j},u_{i})[{\cal X}_{1}(u_{i})a_{\alpha 4}(u,u_{i})\dprod\limits_{k\neq
i}^{3}a_{11}(u_{i},u_{k})+{\cal X}_{2}(u_{i})a_{\alpha
5}(u,u_{i})\dprod\limits_{k\neq i}^{3}a_{21}(u_{i},u_{k})]{\cal B}%
_{3}(u)\Psi _{2}(\overset{\vee }{u_{i}}) 
\]%
\begin{eqnarray}
+ &&\sum_{i=1}^{2}\sum_{j=i+1}^{3}\left\{ {\cal X}_{1}(u_{i}){\cal X}%
_{1}(u_{j})a_{11}(u_{i},u_{m})a_{11}(u_{j},u_{m})H_{\alpha
1}(u_{i},u_{j})\right.  \nonumber \\
&&+\!{\cal X}_{2}(u_{i}){\cal X}%
_{1}(u_{j})a_{21}(u_{i},u_{m})a_{11}(u_{j},u_{m})H_{\alpha 2}(u_{i},u_{j})+%
{\cal X}_{1}(u_{i}){\cal X}%
_{2}(u_{j})a_{11}(u_{i},u_{m})a_{21}(u_{j},u_{m})H_{\alpha 3}(u_{i},u_{j}) 
\nonumber \\
&&\left. +{\cal X}_{2}(u_{i}){\cal X}%
_{2}(u_{j})a_{21}(u_{i},u_{m})a_{21}(u_{j},u_{m})H_{\alpha
4}(u_{i},u_{j})\right\} \dprod\limits_{k=1}^{i-1}\omega
(u_{k},u_{i})\dprod\limits_{l=1\neq i}^{j-1}\omega (u_{l},u_{j}){\cal B}%
_{2}(u)\Psi _{1}(u_{m})  \nonumber \\
&&  \label{baba.65}
\end{eqnarray}%
for $\alpha =1,2,3$ and $m\neq \{i,j\}$. In the above expression we have
introduced the symbol $\overset{\vee }{u_{i}}$ , which as usual means that
the state $\Psi _{2}$ has to be evaluated at spots different form $u_{i}$ .

By performing the action of transfer matrix $t(u)$ on the state $\Psi
_{3}(u_{1},u_{2},u_{3})$, we get%
\begin{equation}
t(u)\Psi _{3}(u_{1},u_{2},u_{3})=\left( \sum_{\alpha =1}^{3}\Omega _{\alpha
}(u){\cal X}_{\alpha }(u)\dprod\limits_{i=1}^{2}a_{\alpha 1}(u,u_{i})\right)
\Psi _{3}(u_{1},u_{2},u_{3})  \label{baba.66}
\end{equation}%
all the unwanted terms vanishing provided that the Bethe equations are
satisfied:%
\begin{equation}
\frac{{\cal X}_{1}(u_{i})}{{\cal X}_{2}(u_{i})}=\Theta
(u_{i})\dprod\limits_{j\neq i=1}^{3}\frac{a_{21}(u_{i},u_{j})}{%
a_{11}(u_{i},u_{j})}\qquad (i=1,2,3)  \label{baba.67}
\end{equation}

\subsection{The n-particle state}

From the previous results one can seek for operator valued functions with a
recurrence relation of the form%
\begin{eqnarray}
\Phi _{n}(u,\ldots ,u_{n}) &=&{\cal B}_{1}(u_{1})\Phi _{n-1}(u_{2},\ldots
,u_{n})  \nonumber \\
&&+{\cal B}_{2}(u_{1})\sum_{i=2}^{n}\digamma _{1}^{(i)}(u_{1},\ldots
,u_{n})\Phi _{n-2}(u_{2},\ldots ,\overset{\vee }{u_{i}},\ldots ,u_{n}){\cal D%
}_{1}(u_{i})  \nonumber \\
&&+{\cal B}_{2}(u_{1})\sum_{i=2}^{n}\digamma _{2}^{(i)}(u_{1},\ldots
,u_{n})\Phi _{n-2}(u_{2},\ldots ,\overset{\vee }{u_{i}},\ldots ,u_{n}){\cal D%
}_{2}(u_{i})  \label{baba.68}
\end{eqnarray}%
It was shown in (\cite{GLL}) that the above operator will be normal ordered
satisfying $n-1$ exchange conditions%
\begin{equation}
\Phi _{n}(u_{1},\ldots ,u_{i},u_{i+1},\ldots ,u_{n})=\omega
(u_{i},u_{i+1})\Phi _{n}(u_{1},\ldots ,u_{i+1},u_{i},\ldots ,u_{n})
\label{baba.69}
\end{equation}%
provided that the functions $\digamma _{\alpha }^{(i)}(u_{1},\ldots ,u_{n})$
are given by 
\begin{equation}
\digamma _{\alpha }^{(i)}(u_{1},\ldots
,u_{n})=\dprod\limits_{j=2}^{i-1}\omega (u_{j},u_{i})\dprod_{k=2,k\neq
i}^{n}a_{\alpha 1}(u_{i},u_{k})G_{d_{\alpha }}(u_{1},u_{i}),\qquad (\alpha
=1,2)  \label{baba.70}
\end{equation}%
Therefore the $n$-particle state will be given by%
\begin{equation}
\Psi _{n}(u_{1},\ldots ,u_{n})=\Phi _{n}(u,\ldots ,u_{n})\left\vert
0\right\rangle   \label{baba.71}
\end{equation}%
and the action of the operators ${\cal D}_{\alpha }(u)$ , $a=1,2,3$ , on
this state will be represented by 
\[
{\cal D}_{\alpha }(u)\Psi _{n}(u_{1},\ldots ,u_{n})={\cal X}_{\alpha
}(u)\dprod\limits_{i=1}^{n}a_{\alpha 1}(u,u_{i})\Psi _{n}(u_{1},\ldots
,u_{n})
\]%
\[
+\sum_{i=1}^{n}\dprod\limits_{j=1}^{i-1}\omega (u_{j},u_{i})[{\cal X}%
_{1}(u_{i})a_{\alpha 2}(u,u_{i})\dprod\limits_{k\neq
i}^{n}a_{11}(u_{i},u_{k})+{\cal X}_{2}(u_{i})a_{\alpha
3}(u,u_{i})\dprod\limits_{k\neq i}^{n}a_{21}(u_{i},u_{k})]{\cal B}%
_{1}(u)\Psi _{n-1}(\overset{\vee }{u_{i}})
\]%
\[
+(1-\delta _{\alpha ,1})\sum_{i=1}^{n}\dprod\limits_{j=1}^{i-1}\omega
(u_{j},u_{i})[{\cal X}_{1}(u_{i})a_{\alpha 4}(u,u_{i})\dprod\limits_{k\neq
i}^{n}a_{11}(u_{i},u_{k})+{\cal X}_{2}(u_{i})a_{\alpha
5}(u,u_{i})\dprod\limits_{k\neq i}^{n}a_{21}(u_{i},u_{k})]{\cal B}%
_{3}(u)\Psi _{n-1}(\overset{\vee }{u_{i}})
\]%
\begin{eqnarray}
&&+\sum_{i=1}^{n-1}\sum_{j=i+1}^{n}\left\{ {\cal X}_{1}(u_{i}){\cal X}%
_{1}(u_{j})\dprod_{k\neq i,j}^{n}a_{11}(u_{i},u_{k})\dprod_{l\neq
i,j}^{n}a_{11}(u_{j},u_{l})H_{\alpha 1}(u_{i},u_{j})\right.   \nonumber \\
&&+{\cal X}_{2}(u_{i}){\cal X}_{1}(u_{j})\dprod_{k\neq
i,j}^{n}a_{21}(u_{i},u_{k})\dprod_{l\neq
i,j}^{n}a_{11}(u_{j},u_{l})H_{\alpha 2}(u_{i},u_{j})  \nonumber \\
&&+{\cal X}_{1}(u_{i}){\cal X}_{2}(u_{j})\dprod_{k\neq
i,j}^{n}a_{11}(u_{i},u_{k})\dprod_{l\neq
i,j}^{n}a_{21}(u_{j},u_{l})H_{\alpha 3}(u_{i},u_{j})  \nonumber \\
&&\left. +{\cal X}_{2}(u_{i}){\cal X}_{2}(u_{j})\dprod_{k\neq
i,j}^{n}a_{21}(u_{i},u_{k})\dprod_{l\neq
i,j}^{n}a_{21}(u_{j},u_{l})H_{\alpha 4}(u_{i},u_{j})\right\}   \nonumber \\
&&\times \dprod\limits_{k=1}^{i-1}\omega (u_{k},u_{i})\dprod\limits_{l=1\neq
i}^{j-1}\omega (u_{l},u_{j}){\cal B}_{2}(u)\Psi _{n-2}(\overset{\vee }{u_{i}}%
,\overset{\vee }{u_{j}})  \label{baba.72}
\end{eqnarray}

Finally, the corresponding n-particle eigenvalue problem will be%
\begin{equation}
t(u)\Psi _{n}(u_{1},\ldots ,u_{n})=\left( \sum_{\alpha =1}^{3}\Omega
_{\alpha }(u){\cal X}_{\alpha }(u)\dprod\limits_{i=1}^{n}a_{\alpha
1}(u,u_{i})\right) \Psi _{n}(u_{1},\ldots ,u_{n})  \label{baba.73}
\end{equation}%
provided that the Bethe equations are satisfied:%
\begin{equation}
\frac{{\cal X}_{1}(u_{k})}{{\cal X}_{2}(u_{k})}=\Theta
(u_{k})\dprod\limits_{j=1,j\neq k}^{n}\frac{a_{21}(u_{k},u_{j})}{%
a_{11}(u_{k},u_{j})},\qquad (k=1,2,\ldots ,n)  \label{baba.74}
\end{equation}

\section{Explicit Solutions}

In this section explicit expressions of the eigenvalue problem are presented
for both models. First we recall the appendix to get the coefficients $%
a_{ij}(u,v)$ which appear effectively in the Bethe Ansatz expressions (\ref%
{baba.73}) and (\ref{baba.74}): 
\begin{eqnarray}
a_{11}(u,v) &=&\frac{x_{1}(v-u)}{x_{2}(v-u)}\frac{x_{2}(u+v)}{x_{1}(u+v)} 
\nonumber \\
a_{21}(u,v) &=&\omega (u,v)[\frac{x_{1}(u+v)x_{4}(u+v)-x_{5}(u+v)y_{5}(u+v)}{%
x_{1}(u+v)x_{2}(u+v)}]  \nonumber \\
a_{31}(u,v) &=&\frac{x_{2}(u-v)}{x_{3}(u-v)}[\frac{%
x_{2}(u+v)^{2}-x_{6}(u+v)y_{6}(u+v)}{x_{2}(u+v)x_{3}(u+v)}]  \label{exsol.1}
\end{eqnarray}%
For the factor with the boundary contributions $\Theta (u_{i})$, we will
consider only the simplest expression%
\begin{equation}
\Theta (u_{i})=-\frac{\Omega _{2}(u)a_{25}(u,u_{i})+\Omega
_{3}(u)a_{35}(u,u_{i})}{\Omega _{2}(u)a_{24}(u,u_{i})+\Omega
_{3}(u)a_{34}(u,u_{i})}  \label{exsol.2}
\end{equation}%
where the coefficients are given by%
\begin{eqnarray}
a_{24}(u,v) &=&[\frac{x_{6}(u-v)}{x3(u-v)}\frac{x_{3}(u+v)}{x_{2}(u+v)}%
-f_{1}(v)\frac{x_{6}(u+v)}{x_{2}(u+v)}]  \nonumber \\
a_{25}(u,v) &=&-\frac{x_{6}(u+v)}{x_{2}(u+v)}  \label{exsol.3}
\end{eqnarray}%
\begin{eqnarray}
a_{34}(u,v) &=&f_{3}(u)[f_{1}(v)\frac{x_{6}(u+v)}{x_{2}(u+v)}-\frac{%
x_{6}(u-v)}{x_{3}(u-v)}\frac{x_{3}(u+v)}{x_{2}(u+v)}]  \nonumber \\
&&+f_{1}(v)\frac{y_{6}(u-v)}{x_{3}(u-v)}[\frac{%
x_{6}(u+v)y_{6}(u+v)-x_{2}^{2}(u+v)}{x_{2}(u+v)x_{3}(u+v)}]  \nonumber \\
&&-[\frac{x_{6}(u-v)y_{6}(u-v)-x_{2}^{2}(u-v)}{x_{3}^{2}(u-v)}]\frac{%
y_{6}(u+v)}{x_{2}(u+v)}  \nonumber \\
a_{35}(u,v) &=&f_{3}(u)\frac{x_{6}(u+v)}{x_{2}(u+v)}+\frac{y_{6}(u-v)}{%
x_{3}(u-v)}[\frac{x_{6}(u+v)y_{6}(u+v)-x_{2}^{2}(u+v)}{x_{2}(u+v)x_{3}(u+v)}]
\label{exsol.4}
\end{eqnarray}

\subsection{ZF Model}

Substituting the matrix elements of the ${\cal R}$ matrix and of the $K$
matrices \ for this model we get the following expressions for the terms
with boundary contributions

\begin{eqnarray}
{\cal X}_{1}(u) &=&-\frac{\beta \sinh u+2\cosh u}{\beta \sinh u-2\cosh u}%
\frac{x_{1}^{2N}(u)}{\rho ^{N}(u)},  \nonumber \\
{\cal X}_{2}(u) &=&\frac{\sinh 2u}{\sinh (2u+2\eta )}\frac{\beta \sinh
(u+2\eta )-2\cosh (u+2\eta )}{\beta \sinh u-2\cosh u}\frac{x_{2}^{2N}(u)}{%
\rho ^{N}(u)},  \nonumber \\
{\cal X}_{3}(u) &=&-\frac{\sinh (2u-\eta )}{\sinh (2u+\eta )}\frac{\beta
\sinh (u+\eta )-2\cosh (u+\eta )}{\beta \sinh (u-\eta )+2\cosh (u-\eta )}%
\frac{\beta \sinh (u+2\eta )-2\cosh (u+2\eta )}{\beta \sinh u-2\cosh u}\frac{%
x_{3}^{2N}(u)}{\rho ^{N}(u)},  \nonumber \\
&&  \label{exsol.5}
\end{eqnarray}%
\begin{eqnarray}
\Omega _{1}(u) &=&-\frac{\sinh (2u+3\eta )}{\sinh (2u+\eta )}\frac{\alpha
\sinh u-2\cosh u}{\alpha \sinh (u+2\eta )-2\cosh (u+2\eta )}\frac{\alpha
\sinh (u+\eta )-2\cosh (u+\eta )}{\alpha \sinh (u+\eta )+2\cosh (u+\eta )}, 
\nonumber \\
\Omega _{2}(u) &=&\frac{\sinh (2u+2\eta )}{\sinh 2u}\frac{\alpha \sinh
u-2\cosh u}{\alpha \sinh (u+2\eta )-2\cosh (u+2\eta )},  \nonumber \\
\Omega _{3}(u) &=&-\frac{\alpha \sinh u+2\cosh u}{\alpha \sinh (u+2\eta
)-2\cosh (u+2\eta )},  \label{exsol.6}
\end{eqnarray}%
and%
\begin{equation}
\Theta (u_{i})=-\frac{\sinh (2u_{i}+2\eta )}{\sinh (2u_{i})}\frac{\alpha
\sinh (u_{i}+\eta )+2\cosh (u_{i}+\eta )}{\alpha \sinh (u_{i}+\eta )+2\cosh
(u_{i}+\eta )}.  \label{exsol.7}
\end{equation}%
The coefficients $a_{i1}(u,v)$ are given by 
\begin{eqnarray}
a_{11}(u,v) &=&\frac{\sinh (u+v)}{\sinh (u+v+2\eta )}\frac{\sinh (u-v-2\eta )%
}{\sinh (u-v)}  \nonumber \\
a_{21}(u,v) &=&\frac{\sinh (u+v)}{\sinh (u+v+2\eta )}\frac{\sinh (u+v+3\eta )%
}{\sinh (u+v+\eta )}\frac{\sinh (u-v+\eta )}{\sinh (u-v-\eta )}\frac{\sinh
(u-v-2\eta )}{\sinh (u-v)}  \nonumber \\
a_{31}(u,v) &=&\frac{\sinh (u-v+\eta )}{\sinh (u-v-\eta )}\frac{\sinh
(u+v+3\eta )}{\sinh (u+v+\eta )}  \nonumber \\
\frac{a_{21}(u,v)}{a_{11}(u,v)} &=&\frac{\sinh (u+v+3\eta )}{\sinh (u+v+\eta
)}\frac{\sinh (u-v+\eta )}{\sinh (u-v-\eta )}  \label{exsol.8}
\end{eqnarray}%
It means that the $n$-particle state $\Psi _{n}(\{u_{i}\})$ (\ref{baba.71})
is an eigenfuction of the {\small ZF} transfer matrix \ with the eigenvalue%
\begin{eqnarray}
\Lambda _{n}(u,\{u_{i}\}) &=&{\cal X}_{1}(u)\Omega
_{1}(u)\dprod\limits_{i=1}^{n}\frac{\sinh (u+u_{i})}{\sinh (u+u_{i}+2\eta )}%
\frac{\sinh (u-u_{i}-2\eta )}{\sinh (u-u_{i})}  \nonumber \\
&&+{\cal X}_{2}(u)\Omega _{2}(u)\dprod\limits_{i=1}^{n}\frac{\sinh (u+u_{i})%
}{\sinh (u+u_{i}+2\eta )}\frac{\sinh (u+u_{i}+3\eta )}{\sinh (u+u_{i}+\eta )}
\nonumber \\
&&\times \dprod\limits_{i=1}^{n}\frac{\sinh (u-u_{i}+\eta )}{\sinh
(u-u_{i}-\eta )}\frac{\sinh (u-u_{i}-2\eta )}{\sinh (u-u_{i})}  \nonumber \\
&&+{\cal X}_{1}(u)\Omega _{1}(u)\dprod\limits_{i=1}^{n}\frac{\sinh
(u-u_{i}+\eta )}{\sinh (u-u_{i}-\eta )}\frac{\sinh (u+u_{i}+3\eta )}{\sinh
(u+u_{i}+\eta )}  \label{exsol.9}
\end{eqnarray}%
provided that the parameters $u_{i}$ satisfy the Bethe equations 
\begin{eqnarray}
\left( \frac{\sinh (u_{i}+2\eta )}{\sinh (u_{i})}\right) ^{2N} &=&\frac{%
\alpha \sinh (u_{i}+\eta )+2\cosh (u_{i}+\eta )}{\alpha \sinh (u_{i}+\eta
)-2\cosh (u_{i}+\eta )}\frac{\beta \sinh (u_{i}+2\eta )-2\cosh (u_{i}+2\eta )%
}{\beta \sinh u_{i}+2\cosh u_{i}}  \nonumber \\
&&\dprod\limits_{j=1,\ j\neq i}^{n}\frac{\sinh (u_{i}+u_{j}+3\eta )}{\sinh
(u_{i}+u_{j}+\eta )}\frac{\sinh (u_{i}-u_{j}+\eta )}{\sinh (u_{i}-u_{j}-\eta
)}  \nonumber \\
i &=&1,2,...,n  \label{exsol.10}
\end{eqnarray}%
Putting these expressions in a symmetric for ($u_{i}\rightarrow u_{i}-\eta $%
) one can see that we have recovered the previous results obtained by the
fusion procedure \cite{Mez, YB}.

\subsection{IK Model}

In this model the $K$ matrices have no free parameters and we have to
consider two cases: \ the quantum group invariant and the non-quantum group
invariant. For both cases the $a_{i1}(u,v)$ coefficients are given by%
\[
a_{11}(u,v)=\frac{\sinh (\frac{1}{2}(u+v))}{\sinh (\frac{1}{2}(u+v)-2\eta )}%
\frac{\sinh (\frac{1}{2}(u-v)+2\eta )}{\sinh (\frac{1}{2}(u-v))}, 
\]%
\[
a_{21}(u,v)=\frac{\sinh (\frac{1}{2}(u+v)-4\eta )}{\sinh (\frac{1}{2}%
(u+v)-2\eta )}\frac{\sinh (\frac{1}{2}(u-v)-2\eta )}{\sinh (\frac{1}{2}(u-v))%
}\frac{\cosh (\frac{1}{2}(u-v)+\eta )}{\cosh (\frac{1}{2}(u-v)-\eta )}\frac{%
\cosh (\frac{1}{2}(u+v)-\eta )}{\cosh (\frac{1}{2}(u+v)-3\eta )}, 
\]%
\[
a_{31}(u,v)=\frac{\cosh (\frac{1}{2}(u-v)-3\eta )}{\cosh (\frac{1}{2}%
(u-v)-\eta )}\frac{\cosh (\frac{1}{2}(u+v)-5\eta )}{\cosh (\frac{1}{2}%
(u+v)-3\eta )}, 
\]%
\begin{equation}
\frac{a_{21}(u,v)}{a_{11}(u,v)}=\frac{\sinh (\frac{1}{2}(u+v)-4\eta )}{\sinh
(\frac{1}{2}(u+v))}\frac{\sinh (\frac{1}{2}(u-v)-2\eta )}{\sinh (\frac{1}{2}%
(u-v)+2\eta )}\frac{\cosh (\frac{1}{2}(u-v)+\eta )}{\cosh (\frac{1}{2}%
(u-v)-\eta )}\frac{\cosh (\frac{1}{2}(u+v)-\eta )}{\cosh (\frac{1}{2}%
(u+v)-3\eta )}.  \label{exsol.11}
\end{equation}%
Though, the boundary contributions are different for each case:

\subsubsection{The quantum group invariant case}

In this case we have%
\begin{eqnarray}
{\cal X}_{1}(u) &=&\frac{x_{1}^{2N}(u)}{\rho ^{N}(u)}  \nonumber \\
{\cal X}_{2}(u) &=&{\rm e}^{2\eta }\frac{\sinh u}{\sinh (u-2\eta )}\frac{%
x_{2}^{2N}(u)}{\rho ^{N}(u)}  \nonumber \\
{\cal X}_{3}(u) &=&{\rm e}^{2\eta }\frac{\sinh u}{\sinh (u-4\eta )}\frac{%
\cosh (u-5\eta )}{\cosh (u-3\eta )}\frac{x_{3}^{2N}(u)}{\rho ^{N}(u)}
\label{exsol.12}
\end{eqnarray}%
\begin{eqnarray}
\Omega _{1}(u) &=&\frac{\sinh (u-6\eta )}{\sinh (u-2\eta )}\frac{\cosh
(u-\eta )}{\cosh (u-3\eta )}  \nonumber \\
\Omega _{2}(u) &=&{\rm e}^{-2\eta }\frac{\sinh (u-6\eta )}{\sinh (u-4\eta )}
\nonumber \\
\Omega _{3}(u) &=&{\rm e}^{-2\eta }  \label{exsol.13}
\end{eqnarray}%
and%
\begin{equation}
\Theta (u_{i})={\rm e}^{-2\eta }\frac{\sinh (u_{i}-2\eta )}{\sinh u_{i}}%
,\qquad i=1,2,...,n  \label{exsol.14}
\end{equation}%
Therefore, the $n$-particle state $\Psi _{n}(\{u_{i}\})$ is an eigenfuction
of the {\small IK} transfer matrix $t(u)$ with eigenvalue%
\begin{eqnarray}
\Lambda _{n}(u,\{u_{i}\}) &=&\frac{\sinh (u-6\eta )}{\sinh (u-2\eta )}\frac{%
\cosh (u-\eta )}{\cosh (u-3\eta )}\frac{x_{1}^{2N}(u)}{\rho ^{N}(u)}%
\dprod\limits_{i=1}^{n}a_{11}(u,u_{i})  \nonumber \\
&&+\frac{\sinh (u-6\eta )}{\sinh (u-4\eta )}\frac{\sinh u}{\sinh (u-2\eta )}%
\frac{x_{2}^{2N}(u)}{\rho ^{N}(u)}\dprod\limits_{i=1}^{n}a_{21}(u,u_{i}) 
\nonumber \\
&&+\frac{\sinh u}{\sinh (u-4\eta )}\frac{\cosh (u-5\eta )}{\cosh (u-3\eta )}%
\frac{x_{3}^{2N}(u)}{\rho ^{N}(u)}\dprod\limits_{i=1}^{n}a_{31}(u,u_{i})
\label{exsol.15}
\end{eqnarray}%
provided that the parameters $u_{i}$ are solutions of the Bethe equations%
\begin{eqnarray}
\left( \frac{\sinh (\frac{1}{2}u_{i}-2\eta )}{\sinh (\frac{1}{2}u_{i})}%
\right) ^{2N} &=&\dprod\limits_{j=1,\ j\neq i}^{n}\frac{\sinh (\frac{1}{2}%
(u_{i}+u_{j})-4\eta )}{\sinh (\frac{1}{2}(u_{i}+u_{j}))}\frac{\sinh (\frac{1%
}{2}(u_{i}-u_{j})-2\eta )}{\sinh (\frac{1}{2}(u_{i}-u_{j})+2\eta )} 
\nonumber \\
&&\times \frac{\cosh (\frac{1}{2}(u_{i}-u_{j})+\eta )}{\cosh (\frac{1}{2}%
(u_{i}-u_{j})-\eta )}\frac{\cosh (\frac{1}{2}(u_{i}+u_{j})-\eta )}{\cosh (%
\frac{1}{2}(u_{i}+u_{j})-3\eta )}  \nonumber \\
i &=&1,2,...,n  \label{exsol.16}
\end{eqnarray}%
Putting these relation in a symmetric form ($u_{i}\rightarrow u_{i}+2\eta $)
one can see that we have the results obtained by (\cite{MN3, YB}) using the
analytical Bethe Ansatz.

\subsubsection{The non-quantum group invariant cases}

These were the cases considered by Guang-Liang Li, {\it et al }\cite{GLL},
where a spurious dependence on the spectral parameter could exist. Here we
shall see that this dependence of the Bethe equations on the spectral
parameter is inexistent..

As the correspondent $K$-matrix solutions are complex conjugated, we will
consider here only the case of $(F^{+},G^{+})$, defined in Section $2$.

A way to turn more direct the comparison of our results with those
previously known is to borrow the notation presented in \cite{YB}: first,
the contributions from the boundaries have the form%
\begin{equation}
{\cal X}_{1}(u)\Omega _{1}(u)=\alpha (u)\frac{x_{1}^{2N}(u)}{\rho ^{N}(u)}%
,\quad {\cal X}_{2}(u)\Omega _{2}(u)=\beta (u)\frac{x_{2}^{2N}(u)}{\rho
^{N}(u)},\quad {\cal X}_{3}(u)\Omega _{3}(u)=\gamma (u)\frac{x_{3}^{2N}(u)}{%
\rho ^{N}(u)}  \label{exsol.17}
\end{equation}%
where%
\begin{eqnarray}
\alpha (u) &=&\frac{\left( f^{+}(u)\right) ^{2}}{x_{1}(2u)}\xi ^{+}(u)\sinh
(u-6\eta )  \nonumber \\
\beta (u) &=&\alpha (u)\frac{\sinh u}{\sinh (u-4\eta )}\frac{\cosh \eta
-i\sinh (u-2\eta )}{\cosh \eta +i\sinh (u-2\eta )}  \nonumber \\
\gamma (u) &=&\alpha (u)\frac{\xi ^{-}(u)}{\xi ^{+}(u)}\frac{\sinh u}{\sinh
(u-4\eta )}\frac{\sinh (u-2\eta )}{\sinh (u-6\eta )}  \label{exsol.18}
\end{eqnarray}%
and%
\begin{equation}
\xi ^{\pm }(u)=2\cosh (u-3\eta )\pm 2i\sinh 2\eta  \label{exsol.19}
\end{equation}%
Second, the boundary factors $\Theta (u_{i})$ are given by%
\begin{equation}
\Theta (u_{i})=\frac{e^{-u_{i}}\sinh (u_{i}-2\eta )}{\sinh u_{i}}\frac{\cosh
(\frac{1}{2}(u+u_{i})-3\eta )+i\sinh (\frac{1}{2}(u-u_{i})-2\eta )}{\cosh (%
\frac{1}{2}(u-u_{i})-\eta )+i\sinh (\frac{1}{2}(u+u_{i})-4\eta )}.
\label{exsol.20a}
\end{equation}%
Here we note that (\ref{exsol.20a}) is the expression obtained in \cite{GLL}
for their $\beta (u,u_{i})$ factors. From this result it follows the
dependence on $u$ in their Bethe equations. However, analyzing the (\ref%
{exsol.20a}) expression carefully, more precisely, using the identity%
\begin{equation}
\frac{\cosh (\frac{1}{2}(u+u_{i})-3\eta )+i\sinh (\frac{1}{2}(u-u_{i})-2\eta
)}{\cosh (\frac{1}{2}(u-u_{i})-\eta )+i\sinh (\frac{1}{2}(u+u_{i})-4\eta )}=%
\frac{\cosh \eta -i\sinh (u_{i}-2\eta )}{\cosh (u_{i}-3\eta )}
\label{exsol.20b}
\end{equation}%
one we can see that this $u$ dependency is only apparent. Therefore the
correct expressions for \ $\Theta (u_{i})$ are given by 
\begin{equation}
\Theta (u_{i})={\rm e}^{-u_{i}}\frac{\sinh (u_{i}-2\eta )}{\sinh (u_{i})}%
\frac{\cosh \eta -i\sinh (u_{i}-2\eta )}{\cosh (u_{i}-3\eta )},\quad
i=1,...,n.  \label{exsol.20c}
\end{equation}%
Next, to recover our notation we will need of the following identity%
\begin{equation}
\frac{\Omega _{2}(u)}{\Omega _{1}(u)}=\Theta (u)\frac{\sinh u}{\sinh
(u-4\eta )}\frac{\cosh (u-3\eta )}{\cosh (u-\eta )}  \label{exsol.21}
\end{equation}

Therefore the $n$-particle state $\Psi _{n}(\{u_{i}\})$ is an eigenfunction
of the {\small IK} transfer matrix with the boundary contribution ($%
F^{+},G^{+}$), and has eigenvalue%
\begin{equation}
\Lambda _{n}(u,\{u_{i}\})=\alpha (u)\frac{x_{1}^{2N}(u)}{\rho ^{N}(u)}%
\dprod\limits_{i=1}^{n}a_{11}(u,u_{i})+\beta (u)\frac{x_{1}^{2N}(u)}{\rho
^{N}(u)}\dprod\limits_{i=1}^{n}a_{21}(u,u_{i})+\gamma (u)\frac{x_{1}^{2N}(u)%
}{\rho ^{N}(u)}\dprod\limits_{i=1}^{n}a_{31}(u,u_{i})  \label{exsol.22}
\end{equation}%
provided that the $u_{i}$ parameters satisfy the Bethe equations%
\begin{eqnarray}
\left( \frac{\sinh (\frac{1}{2}u_{i}-2\eta )}{\sinh (\frac{1}{2}u_{i})}%
\right) ^{2N} &=&\frac{\beta (u_{i})}{\alpha (u_{i})}\frac{\sinh
(u_{i}-4\eta )}{\sinh u_{i}}\frac{\cosh (u_{i}-\eta )}{\cosh (u_{i}-3\eta )}
\nonumber \\
&&\times \dprod\limits_{j=1,\ j\neq i}^{n}\frac{\sinh (\frac{1}{2}%
(u_{i}+u_{j})-4\eta )}{\sinh (\frac{1}{2}(u_{i}+u_{j}))}\frac{\sinh (\frac{1%
}{2}(u_{i}-u_{j})-2\eta )}{\sinh (\frac{1}{2}(u_{i}-u_{j})+2\eta )} 
\nonumber \\
&&\times \frac{\cosh (\frac{1}{2}(u_{i}-u_{j})+\eta )}{\cosh (\frac{1}{2}%
(u_{i}-u_{j})-\eta )}\frac{\cosh (\frac{1}{2}(u_{i}+u_{j})-\eta )}{\cosh (%
\frac{1}{2}(u_{i}+u_{j})-3\eta )}  \label{exsol.23}
\end{eqnarray}%
Writing the above Bethe equations in a symmetric form ($u_{i}\rightarrow
u_{i}-\eta $) one can see that is exactly the result obtained in \cite{YB}
using the analytical {\small BA}.

\section{Conclusion}

The main result of this paper may be summarized \ by saying that the
Sklyanin algebraic Bethe Ansatz is unique and consistent for the {\small ZF}
and the {\small IK} vertex models with boundaries, corroborating all
previous investigations in these models. Thus the Bethe vectors here
obtained can be used for instance to obtain the eingenvalue problem for the
corresponding Gaudin models \cite{GAU} with boundaries and their associated
systems of Knizhnik-Zamolodchikov equations \cite{KZ}. Also, our results
encourage the believe that other $19$-vertex models , such as the $osp(2|1)$
model and the $sl(2|1)^{(2)}$ model in the presence of boundaries can also
be solved by the algebraic Bethe Ansatz. In fact, preliminary results \cite%
{KLS} indicate that this is indeed true.

It was in the {\small IK} model that Tarasov developed the algebraic {\small %
BA} for the three-state models with periodic boundary conditions \cite{TA}.
Here, with the aid of previous works \cite{FAN, GLL}, two of the three-state 
$19$-vertex models have their algebraic Bethe Ansatz derived using a
generalization of the Tarasov's approach. The algebraic {\small BA} for $n$%
-state models with periodic boundary conditions was developed by Martins in 
\cite{MAR}.Therefore we believe that the Martins's approach can be
generalized to include the diagonal open boundary conditions.

{\bf Acknowledgment:} This work was supported in part by Funda\c{c}\~{a}o de
Amparo \`{a} Pesquisa do Estado de S\~{a}o Paulo-{\small FAPESP}-Brasil and
by Conselho Nacional de Desenvolvimento-{\small CNPq}-Brasil.

\newpage

\appendix{}

\section{The Commutation Relations}

The equation (\ref{baba.26}) gives us the commutation relations for the
matrix elements of the double-row monodromy matrix which play a fundamental
role in the algebraic Bethe Ansatz. Here we present the commutations
relations and their coefficientes using a compact notation. Recall that the
operator 
\begin{equation}
{\cal U}(u)=\left( 
\begin{array}{ccc}
{\cal D}_{1}(u) & {\cal B}_{1}(u) & {\cal B}_{2}(u) \\ 
{\cal C}_{1}(u) & {\cal D}_{2}(u) & {\cal B}_{3}(u) \\ 
{\cal C}_{2}(u) & {\cal C}_{3}(u) & {\cal D}_{3}(u)%
\end{array}%
\right) ,  \label{A.1}
\end{equation}%
satisfy the fundamental reflection equation%
\begin{equation}
{\cal R}_{12}(u-v){\cal U}_{1}(u){\cal R}_{21}(u+v){\cal U}_{2}(v)={\cal U}%
_{2}(v){\cal R}_{12}(u+v){\cal U}_{1}(u){\cal R}_{21}(u-v).  \label{A.2}
\end{equation}%
Substituting (\ref{des.18}) and (\ref{A.1}) into (\ref{A.2}) we get $81$
equations involving products of two matrix elements of ${\cal U}(u)$. These
equations can be manipulated in ordem to put the product of pairs of
operators in the normal ordered form. To do this we shall proceed in the
following way. First we denote by $E[i,j]=0$ the $(i,j)$ component of the
matrix equation (\ref{A.2}) and collect them in blocks $%
B[i,j],i=1,...,5,j=i,...,10-i,$ defined by%
\begin{equation}
B[i,j]=\left\{ F_{ij}=E[i,j],\qquad f_{ij}=E[j,i],\qquad
FF_{ij}=E[10-i,10-j],\qquad ff_{ij}=E[10-j,10-i]\right\}  \label{A.3}
\end{equation}%
\TEXTsymbol{>}From these blocks we can see that the pair $(F_{ij},f_{ij})$
as well as $(FF_{ij},ff_{ij})$ \ can be solved simultaneously.

We introduce the notation \ 
\begin{equation}
D_{i}={\cal D}_{i}(u),\quad d_{i}={\cal D}_{i}(v),\quad B_{i}={\cal B}%
_{i}(u),\quad b_{i}={\cal B}_{i}(v),\quad C_{i}={\cal C}_{i}(u),\quad c_{i}=%
{\cal C}_{i}(v)  \label{A.4}
\end{equation}%
for the operators of the double-row monodromy matrix and%
\begin{equation}
X_{i}=x_{i}(u+v),\quad Y_{i}=y_{i}(u+v),\quad x_{i}=x_{i}(u-v),\quad
y_{i}=y_{i}(u-v),  \label{A.5}
\end{equation}%
for the Boltzmann weigths and%
\begin{equation}
\{Z\}_{ij}=\{z\}_{ij}(v,u),\qquad \{z\}_{ij}=\{z\}_{ij}(u,v),\quad
F_{i}=f_{i}(u),\quad f_{i}=f_{i}(v).  \label{A.6}
\end{equation}%
for the coefficientes of the commutations relations, where 
\begin{equation}
\{Z\}=A,B,C,D,E,X,Y\qquad {\rm and}\qquad \{z\}=a,b,c,d,e,x,y.
\end{equation}

Taking into account these simplifications, we will indicate the pair $%
(F_{ij},f_{ij})$ or $(FF_{ij},ff_{ij})$ for which the corresponding normal
ordered relations were obtained:

\begin{itemize}
\item $(F_{14},f_{14})$%
\begin{eqnarray}
D_{1}b_{1}
&=&a_{11}b_{1}D_{1}+a_{12}B_{1}d_{1}+a_{13}B_{1}d_{2}+a_{14}B_{2}c_{1}+a_{15}B_{2}c_{3}+a_{16}b_{2}C_{1}
\nonumber \\
C_{1}d_{1}
&=&A_{11}d_{1}C_{1}+A_{12}D_{1}c_{1}+A_{13}D_{2}c_{1}+A_{14}B_{1}c_{2}+A_{15}B_{3}c_{2}+A_{16}b_{1}C_{2}
\label{A.7}
\end{eqnarray}%
where the coefficients are%
\begin{eqnarray}
A_{11} &=&\frac{x_{1}}{x_{2}}\frac{X_{2}}{X_{1}},\quad A_{12}=-F_{1}\frac{%
X_{5}}{X_{1}}-\frac{y_{5}}{x_{2}}\frac{X_{2}}{X_{1}},\quad A_{13}=-\frac{%
X_{5}}{X_{1}},\quad  \nonumber \\
A_{14} &=&-\frac{y_{5}}{x_{2}}\frac{X_{6}}{X_{1}},\quad A_{15}=-\frac{X_{7}}{%
X_{1}},\quad A_{16}=\frac{x_{1}}{x_{2}}\frac{X_{6}}{X_{1}}.  \label{A.8}
\end{eqnarray}

\item $(F_{17},f_{17})$ 
\begin{eqnarray}
D_{1}b_{2}
&=&b_{11}b_{2}D_{1}+b_{12}B_{2}d_{1}+b_{13}B_{2}d_{2}+b_{14}B_{2}d_{3}+b_{15}B_{1}b_{1}+b_{16}B_{1}b_{3}
\nonumber \\
C_{2}d_{1}
&=&B_{11}d_{1}C_{2}+B_{12}D_{1}c_{2}+B_{13}D_{2}c_{2}+B_{14}D_{3}c_{2}+B_{15}C_{1}c_{1}+B_{16}C_{3}c_{1}
\label{A.9}
\end{eqnarray}%
where%
\begin{eqnarray}
B_{11} &=&\frac{x_{1}}{x_{3}}\frac{X_{3}}{X_{1}},\quad B_{12}=-F_{1}\frac{%
y_{6}}{x_{3}}\frac{X_{6}}{X_{1}}-F_{2}\frac{X_{7}}{X_{1}}-\frac{y_{7}}{x_{3}}%
\frac{X_{3}}{X_{1}},\quad B_{13}=-\frac{y_{6}}{x_{3}}\frac{X_{6}}{X_{1}}%
-F_{3}\frac{X_{7}}{X_{1}},\quad  \nonumber \\
B_{14} &=&-\frac{X_{7}}{X_{1}},\quad B_{15}=-\frac{y_{6}}{x_{3}}\frac{X_{2}}{%
X_{1}},\quad B_{16}=-\frac{X_{5}}{X_{1}}.  \label{A.10}
\end{eqnarray}

\item $(FF_{36},ff_{36})$ 
\begin{eqnarray}
D_{1}b_{3}
&=&x_{11}b_{3}D_{1}+x_{12}b_{1}D_{1}+x_{13}B_{1}d_{1}+x_{14}B_{1}d_{2}+x_{15}B_{1}d_{3}+x_{16}B_{2}c_{1}
\nonumber \\
&&+x_{17}B_{2}c_{3}+x_{18}b_{2}C_{1}  \nonumber \\
C_{3}d_{1}
&=&X_{11}d_{1}C_{3}+X_{12}d_{1}C_{1}+X_{13}D_{1}c_{1}+X_{14}D_{2}c_{1}+X_{15}D_{3}c_{1}+X_{16}B_{1}c_{2}
\nonumber \\
&&+X_{17}B_{3}c_{2}+X_{18}b_{1}C_{2}  \label{A.11}
\end{eqnarray}%
with the following coefficients%
\begin{eqnarray}
X_{11} &=&\frac{x_{2}}{x_{3}}\frac{X_{3}}{X_{2}},\quad X_{12}=\frac{y_{5}}{%
x_{3}}\frac{Y_{6}}{X_{2}}-\frac{y_{6}}{x_{3}}\frac{Y_{5}}{X_{2}}A_{11},\quad
\nonumber \\
X_{13} &=&-F_{1}\frac{y_{6}}{x_{3}}\frac{X_{4}}{X_{2}}-F_{2}\frac{X_{6}}{%
X_{2}}-\frac{y_{7}}{x_{3}}\frac{Y_{6}}{X_{2}}-\frac{y_{6}}{x_{3}}\frac{Y_{5}%
}{X_{2}}A_{12}  \nonumber \\
X_{14} &=&-F_{3}\frac{X_{6}}{X_{2}}-\frac{y_{6}}{x_{3}}\frac{X_{4}}{X_{2}}-%
\frac{y_{6}}{x_{3}}\frac{Y_{5}}{X_{2}}A_{13},\quad  \nonumber \\
X_{15} &=&-\frac{X_{6}}{X_{2}},\quad X_{16}=-\frac{y_{7}}{x_{3}}-\frac{y_{6}%
}{x_{3}}\frac{Y_{5}}{X_{2}}A_{14}  \nonumber \\
X_{17} &=&-\frac{y_{6}}{x_{3}}\frac{X_{5}}{X_{2}}-\frac{y_{6}}{x_{3}}\frac{%
Y_{5}}{X_{2}}A_{15},\quad X_{18}=\frac{y_{5}}{x_{3}}-\frac{y_{6}}{x_{3}}%
\frac{Y_{5}}{X_{2}}A_{16}.  \label{A.12}
\end{eqnarray}%
Note that for each pair of equations the corresponding commutation relations
are related by interchanging 
\begin{equation}
u\leftrightarrow v,\qquad D_{i}\leftrightarrow d_{i},\qquad
B_{i}\leftrightarrow c_{i},\qquad C_{i}\leftrightarrow b_{i}  \label{A.13}
\end{equation}

\item $(F_{24},f_{24})$ 
\begin{eqnarray}
C_{1}b_{1}
&=&c_{11}b_{1}C_{1}+c_{12}b_{1}C_{3}+c_{13}B_{1}c_{3}+c_{14}B_{3}c_{3}+c_{15}b_{2}C_{2}+c_{16}d_{1}D_{1}+c_{17}d_{1}D_{2}
\nonumber \\
&&+c_{18}D_{1}d_{1}+c_{19}D_{1}d_{2}+c_{110}D_{2}d_{1}+c_{111}D_{2}d_{2}
\label{A.14}
\end{eqnarray}%
with the following coefficients%
\begin{eqnarray}
c_{11} &=&\frac{X_{4}}{X_{1}},\quad c_{12}=\frac{x_{5}}{x_{2}}\frac{X_{6}}{%
X_{1}},\quad c_{13}=-\frac{y_{5}}{x_{2}}\frac{X_{6}}{X_{1}},\quad c_{14}=-%
\frac{X_{7}}{X_{1}},\quad c_{15}=\frac{X_{5}}{X_{1}},  \nonumber \\
c_{16} &=&\frac{Y_{5}}{X_{1}}+F_{1}\frac{x_{5}}{x_{2}}\frac{X_{2}}{X_{1}}%
,\quad c_{17}=\frac{x_{5}}{x_{2}}\frac{X_{2}}{X_{1}},\quad c_{18}=-f_{1}(%
\frac{y_{5}}{x_{2}}\frac{X_{2}}{X_{1}}+F_{1}\frac{X_{5}}{X_{1}}),  \nonumber
\\
c_{19} &=&-(\frac{y_{5}}{x_{2}}\frac{X_{2}}{X_{1}}+F_{1}\frac{X_{5}}{X_{1}}%
),\quad c_{110}=-f_{1}\frac{X_{5}}{X_{1}},\quad c_{111}=-\frac{X_{5}}{X_{1}}.
\label{A.15}
\end{eqnarray}%
The pairs $(F_{16},f_{16}),\ (F_{18},f_{18}),\ (FF_{16},ff_{16})$, $%
(FF_{18},ff_{18})$ and $F_{19}$ form a closed set of equations from which we
have derived the following commutation relations%
\begin{eqnarray}
B_{2}b_{1} &=&e_{11}b_{1}B_{2}+e_{12}b_{2}B_{1}+e_{13}b_{2}B_{3},  \nonumber
\\
C_{1}c_{2} &=&E_{11}c_{2}C_{1}+E_{12}c_{1}C_{2}+E_{13}c_{3}C_{2}  \nonumber
\\
B_{3}b_{2} &=&e_{41}b_{2}B_{3}+e_{42}b_{1}B_{2}+e_{43}b_{3}B_{2}  \nonumber
\\
C_{2}c_{3} &=&E_{41}c_{3}C_{2}+E_{42}c_{2}C_{1}+E_{43}c_{2}C_{3}  \nonumber
\\
B_{1}b_{2}
&=&e_{21}b_{2}B_{1}+e_{22}b_{2}B_{3}+e_{23}b_{1}B_{2}+e_{24}b_{3}B_{2} 
\nonumber \\
C_{2}c_{1}
&=&E_{21}c_{1}C_{2}+E_{22}c_{3}C_{2}+E_{23}c_{2}C_{1}+E_{24}c_{2}C3 
\nonumber \\
B_{2}b_{3}
&=&e_{31}b_{3}B_{2}+e_{32}b_{1}B_{2}+e_{33}b_{2}B_{1}+e_{34}b_{2}B_{3} 
\nonumber \\
C_{3}c_{2}
&=&E_{31}c_{2}C_{3}+E_{32}c_{2}C_{1}+E_{33}c_{1}C_{2}+E_{34}c_{3}C_{2} 
\nonumber \\
B_{2}b_{2} &=&b_{2}B_{2},\qquad C_{2}c_{2}=c_{2}C_{2}  \label{A.16}
\end{eqnarray}%
where%
\begin{equation}
e_{11}=\frac{x_{2}}{x_{1}}\frac{X_{2}}{X_{3}},\quad e_{12}=\frac{y_{5}}{x_{1}%
},\quad e_{13}=\frac{x_{2}}{x_{1}}\frac{X_{6}}{X_{3}}.
\end{equation}%
\begin{equation}
E_{41}=\frac{x_{2}}{x_{1}}\frac{X_{2}}{X_{3}},\quad E_{42}=\frac{x_{2}}{x_{1}%
}\frac{Y_{6}}{X_{3}},\quad E_{43}=\frac{x_{5}}{x_{1}}
\end{equation}%
\begin{eqnarray}
e_{21} &=&\frac{x_{2}}{x_{1}}\frac{X_{2}X_{3}}{X_{2}^{2}-X_{6}Y_{6}},\quad
e_{22}=\frac{x_{5}(x_{1}^{2}+x_{2}^{2}-x_{5}y_{5})}{%
x_{1}(x_{2}^{2}-x_{5}y_{5})}\frac{X_{2}X_{6}}{X_{2}^{2}-X_{6}Y_{6}} 
\nonumber \\
e_{23} &=&\frac{x_{1}x_{5}}{x_{2}^{2}-x_{5}y_{5}}\frac{X_{6}Y_{6}}{%
X_{2}^{2}-X_{6}Y_{6}}+\frac{x_{5}}{x_{1}}\frac{X_{2}^{2}}{%
X_{2}^{2}-X_{6}Y_{6}},\quad e_{24}=-\frac{x_{1}x_{2}}{x_{2}^{2}-x_{5}y_{5}}%
\frac{X_{3}X_{6}}{X_{2}^{2}-X_{6}Y_{6}}
\end{eqnarray}%
\begin{eqnarray}
e_{31} &=&\frac{x_{1}x_{2}}{x_{2}^{2}-x_{5}y_{5}}\frac{X_{2}X_{3}}{%
X_{2}^{2}-X_{6}Y_{6}},\quad e_{32}=-\frac{%
x_{5}(x_{1}^{2}+x_{2}^{2}-x_{5}y_{5})}{x_{1}(x_{2}^{2}-x_{5}y_{5})}\frac{%
X_{2}Y_{6}}{X_{2}^{2}-X_{6}Y_{6}}  \nonumber \\
e_{33} &=&-\frac{x_{2}}{x_{1}}\frac{X_{3}Y_{6}}{X_{2}^{2}-X_{6}Y_{6}},\quad
e_{34}=\frac{x_{5}}{x_{1}}\frac{X_{6}Y_{6}}{X_{2}^{2}-X_{6}Y_{6}}-\frac{%
x_{1}x_{5}}{x_{2}^{2}-x_{5}y_{5}}\frac{X_{2}^{2}}{X_{2}^{2}-X_{6}Y_{6}}
\end{eqnarray}

\item $(F_{45},f_{45})$
\end{itemize}

\begin{eqnarray}
D_{2}b_{1}
&=&a_{21}b_{1}D_{2}+a_{22}B_{1}d_{1}+a_{23}B_{1}d_{2}+a_{24}B_{3}d_{1}+a_{25}B_{3}d_{2}+a_{26}B_{2}c_{1}
\nonumber \\
&&+a_{27}B_{2}c_{3}+a_{28}b_{2}C_{1}+a_{29}b_{2}C_{3}  \nonumber \\
C_{1}d_{2}
&=&A_{21}d_{2}C_{1}+A_{22}D_{1}c_{1}+A_{23}D_{2}c_{1}+A_{24}D_{1}c_{3}+A_{25}D_{2}c_{3}+A_{26}B_{1}c_{2}
\nonumber \\
&&+A_{27}B_{3}c_{2}+A_{28}b_{1}C_{2}+A_{29}b_{3}C_{2}
\end{eqnarray}

with%
\begin{eqnarray}
a_{21} &=&\frac{x_{4}}{x_{2}}\frac{X_{4}}{X_{2}}+F_{3}\frac{x_{6}}{x_{2}}%
\frac{X_{6}}{X_{2}}+\frac{x_{4}}{x_{2}}\frac{Y_{5}}{X_{2}}A_{13}+\frac{x_{6}%
}{x_{2}}X_{14}  \nonumber \\
a_{22} &=&-f_{1}\frac{y_{5}}{x_{2}}\frac{X_{4}}{X_{2}}-(F_{1}+\frac{y_{5}}{%
x_{2}}\frac{Y_{5}}{X_{2}})a_{12}+\frac{x_{4}}{x_{2}}\frac{Y_{5}}{X_{2}}%
A_{11}+\frac{x_{6}}{x_{2}}X_{12}  \nonumber \\
a_{23} &=&-\frac{y_{5}}{x_{2}}\frac{X_{4}}{X_{2}}-(F_{1}+\frac{y_{5}}{x_{2}}%
\frac{Y_{5}}{X_{2}})a_{13},\quad a_{24}=-f_{1}\frac{X_{6}}{X_{2}}+\frac{x_{6}%
}{x_{2}}X_{11},\quad a_{25}=-\frac{X_{6}}{X_{2}}  \nonumber \\
a_{26} &=&-(F_{1}+\frac{y_{5}}{x_{2}}\frac{Y_{5}}{X_{2}})a_{14}+\frac{x_{4}}{%
x_{2}}\frac{Y_{5}}{X_{2}}A_{16}+\frac{x_{6}}{x_{2}}X_{18}  \nonumber \\
a_{27} &=&-\frac{y_{5}}{x_{2}}\frac{X_{5}}{X_{2}}-(F_{1}+\frac{y_{5}}{x_{2}}%
\frac{Y_{5}}{X_{2}})a_{15},\quad a_{28}=\frac{y_{6}}{x_{2}}-(F_{1}+\frac{%
y_{5}}{x_{2}}\frac{Y_{5}}{X_{2}})a_{16}+\frac{x_{4}}{x_{2}}\frac{Y_{5}}{X_{6}%
}A_{14}+\frac{x_{6}}{x_{2}}X_{16}  \nonumber \\
a_{29} &=&\frac{x_{4}}{x_{2}}\frac{X_{5}}{X_{6}}+\frac{x_{4}}{x_{2}}\frac{%
Y_{5}}{X_{6}}A_{15}+\frac{x_{6}}{x_{2}}X_{17}
\end{eqnarray}

\begin{itemize}
\item $(F_{15},f_{15})$ 
\begin{eqnarray}
B_{1}b_{1}
&=&e_{01}b_{1}B_{1}+e_{02}b_{2}D_{2}+e_{03}b_{2}D_{1}+e_{04}B_{2}d_{1}+e_{05}B_{2}d_{2}
\nonumber \\
C_{1}c_{1}
&=&E_{01}c_{1}C_{1}+E_{02}d_{2}C_{2}+E_{03}d_{1}C_{2}+E_{04}D_{1}c_{2}+E_{05}D_{2}c_{2}
\end{eqnarray}%
with%
\begin{eqnarray}
e_{01} &=&\frac{x_{3}x_{4}-x_{6}y_{6}}{x_{1}x_{3}},\quad e_{02}=\frac{%
x_{3}x_{4}-x_{6}y_{6}}{x_{1}x_{3}}\frac{X_{6}}{X_{2}}  \nonumber \\
e_{03} &=&\frac{x_{3}y_{6}-x_{6}y_{7}}{x_{1}x_{3}}\frac{X_{3}}{X_{2}}+F_{1}%
\frac{x_{3}x_{4}-x_{6}y_{6}}{x_{1}x_{3}}\frac{X_{6}}{X_{2}}  \nonumber \\
e_{04} &=&\frac{x_{6}}{x_{3}}\frac{X_{3}}{X_{2}}-f_{1}\frac{X_{6}}{X_{2}}%
,\quad e_{05}=-\frac{X_{6}}{X_{2}}
\end{eqnarray}%
These commutaion relation play a special role in the construction of the $n$%
-particle states, as mentioned above.

\item $(FF_{26},ff_{26})$ 
\begin{eqnarray}
B_{1}b_{3}
&=&d_{11}b_{3}B_{1}+d_{12}b_{1}B_{1}+d_{13}b_{2}D_{1}+d_{14}b_{2}D_{2}+b_{15}B_{2}d_{1}+d_{16}B_{2}d_{2}+d_{17}B_{2}d_{3}
\nonumber \\
C_{3}c_{1}
&=&D_{11}c_{1}C_{3}+D_{12}c_{1}C_{1}+D_{13}d_{1}C_{2}+D_{14}d_{2}C_{2}+D_{15}D_{1}c_{2}+D_{16}D_{2}c_{2}+D_{17}D_{3}c_{2}
\nonumber \\
&&
\end{eqnarray}%
where%
\begin{eqnarray}
D_{11} &=&\frac{X_{3}}{X_{4}+Y_{5}B_{16}},\quad D_{12}=-\frac{%
(y_{5}Y_{6}+x_{2}Y_{5}B_{15})E_{01}-y_{5}Y_{6}}{x_{2}(X_{4}+Y_{5}B_{16})} 
\nonumber \\
D_{13} &=&-\frac{%
(y_{5}Y_{6}+x_{2}Y_{5}B_{15})E_{03}-f_{1}y_{5}X_{2}+x_{2}Y_{5}B_{11}}{%
x_{2}(X_{4}+Y_{5}B_{16})},\quad  \nonumber \\
D_{14} &=&-\frac{(y_{5}Y_{6}+x_{2}Y_{5}B_{15})E_{02}-y_{5}X_{2}}{%
x_{2}(X_{4}+Y_{5}B_{16})}  \nonumber \\
D_{15} &=&-\frac{%
(y_{5}Y_{6}+x_{2}Y_{5}B_{15})E_{04}+F_{1}y_{5}X_{2}+F_{2}x_{2}X_{5}+x_{2}Y_{5}B_{12}%
}{x_{2}(X_{4}+Y_{5}B_{16})}  \nonumber \\
D_{16} &=&-\frac{%
(y_{5}Y_{6}+x_{2}Y_{5}B_{15})E_{05}+y_{5}X_{2}+F_{3}x_{2}X_{5}+x_{2}Y_{5}B_{13}%
}{x_{2}(X_{4}+Y_{5}B_{16})},  \nonumber \\
D_{17} &=&-\frac{X_{5}+Y_{5}B_{14}}{X_{4}+Y_{5}B_{16}}
\end{eqnarray}
\end{itemize}

and%
\begin{eqnarray}
B_{3}b_{1}
&=&d_{21}b_{1}B_{3}+d_{22}b_{1}B_{1}+d_{23}b_{2}D_{1}+d_{24}b_{2}D_{2}+d_{25}b_{2}D_{3}+d_{26}B_{2}d_{1}+d_{27}B_{2}d_{2}
\nonumber \\
C_{1}c_{3}
&=&D_{21}c_{3}C_{1}+D_{22}c_{1}C_{1}+D_{23}d_{1}C_{2}+D_{24}d_{2}C_{2}+D_{25}d_{3}C_{2}+D_{26}D_{1}c_{2}+D_{27}D_{2}c_{2}
\nonumber \\
&&
\end{eqnarray}%
where%
\begin{eqnarray}
d_{21} &=&\frac{X_{4}}{X_{3}}+\frac{Y_{5}}{X_{3}}B_{16},\quad d_{22}=\frac{%
y_{5}}{x_{2}}\frac{Y_{6}}{X_{3}}-\frac{y_{5}}{x_{2}}\frac{Y_{6}}{X_{3}}%
e_{01}+\frac{Y_{5}}{X_{3}}B_{15}  \nonumber \\
d_{23} &=&F_{1}\frac{y_{5}}{x_{2}}\frac{X_{2}}{X_{3}}+F_{2}\frac{X_{5}}{X_{3}%
}-\frac{y_{5}}{x_{2}}\frac{Y_{6}}{X_{3}}e_{03}+\frac{Y_{5}}{X_{3}}B_{12} 
\nonumber \\
d_{24} &=&\frac{y_{5}}{x_{2}}\frac{X_{2}}{X_{3}}+F_{3}\frac{X_{5}}{X_{3}}-%
\frac{y_{5}}{x_{2}}\frac{Y_{6}}{X_{3}}e_{02}+\frac{Y_{5}}{X_{3}}B_{13},\quad
d_{25}=\frac{X_{5}}{X_{3}}+\frac{Y_{5}}{X_{3}}B_{14}  \nonumber \\
d_{26} &=&f_{1}\frac{y_{5}}{x_{2}}\frac{X_{2}}{X_{3}}-\frac{y_{5}}{x_{2}}%
\frac{Y_{6}}{X_{3}}e_{04}+\frac{Y_{5}}{X_{3}}B_{11},\quad  \nonumber \\
d_{27} &=&-\frac{y_{5}}{x_{2}}\frac{X_{2}}{X_{3}}-\frac{y_{5}}{x_{2}}\frac{%
Y_{6}}{X_{3}}e_{05}
\end{eqnarray}

\begin{itemize}
\item $(F_{28},f_{28})$%
\begin{eqnarray}
D_{2}b_{2}
&=&b_{21}b_{2}D_{2}+b_{22}B_{2}d_{1}+b_{23}B_{2}d_{2}+b_{24}B_{2}d_{3}+b_{25}B_{1}b_{1}+b_{26}B_{1}b_{3}
\nonumber \\
&&+b_{27}B_{3}b_{1}+b_{28}B_{3}b_{3}  \nonumber \\
C_{2}d_{2}
&=&B_{21}d_{2}C_{2}+B_{22}D_{1}c_{2}+B_{23}D_{2}c_{2}+B_{24}D_{3}c_{2}+B_{25}C_{1}c_{1}+B_{26}C_{3}c_{1}
\nonumber \\
&&+B_{27}C_{1}c_{3}+B_{28}C_{3}c_{3}
\end{eqnarray}%
where%
\begin{eqnarray}
B_{21} &=&1+\frac{x_{5}}{x_{2}}\frac{X_{3}}{X_{2}}d_{27}+\frac{Y_{6}}{X_{2}}%
e_{05},\quad B_{22}=-f_{1}B_{12}+\frac{x_{5}}{x_{2}}\frac{X_{3}}{X_{2}}%
d_{23}+\frac{Y_{6}}{X_{2}}e_{03}  \nonumber \\
B_{23} &=&-f_{1}B_{13}+\frac{x_{5}}{x_{2}}\frac{X_{3}}{X_{2}}d_{24}+\frac{%
Y_{6}}{X_{2}}e_{02},\quad B_{24}=-f_{1}B_{14}+\frac{x_{5}}{x_{2}}\frac{X_{3}%
}{X_{2}}d_{25}  \nonumber \\
B_{25} &=&-f_{1}B_{15}+\frac{x_{5}}{x_{2}}\frac{X_{3}}{X_{2}}d_{22}+\frac{%
Y_{6}}{X_{2}}e_{01},\quad B_{26}=-f_{1}B_{16}+\frac{x_{5}}{x_{2}}\frac{X_{3}%
}{X_{2}}d_{21}  \nonumber \\
B_{27} &=&-\frac{y_{5}}{x_{2}}\frac{X_{3}}{X_{2}},\quad B_{28}=-\frac{X_{6}}{%
X_{2}}
\end{eqnarray}

\item $(FF_{35},ff_{35})$%
\begin{eqnarray}
C_{3}b_{1}
&=&c_{21}b_{1}C_{1}+c_{22}b_{1}C_{3}+c_{23}B_{1}c_{3}+c_{24}B_{3}c_{3}+c_{25}b_{2}C_{2}+c_{26}d_{1}D_{1}
\nonumber \\
&&+c_{27}d_{1}D_{2}+c_{28}d_{1}D_{3}+c_{29}D_{1}d_{1}+c_{210}D_{1}d_{2}+c_{211}D_{2}d_{1}+c_{212}D_{2}d_{2}
\nonumber \\
&&+c_{213}D_{3}d_{1}+c_{214}D_{3}d_{2}  \nonumber \\
C_{1}b_{3}
&=&C_{21}b_{1}C_{1}+C_{22}b_{3}C_{1}+C_{23}B_{3}c_{1}+C_{24}B_{3}c_{3}+C_{25}b_{2}C_{2}+C_{26}d_{1}D_{1}
\nonumber \\
&&+C_{27}d_{2}D_{1}+C_{28}d_{3}D_{1}+C_{29}D_{1}d_{1}+C_{210}D_{2}d_{1}+C_{211}D_{1}d_{2}
\nonumber \\
&&+C_{213}D_{1}d_{3}+C_{214}D_{2}d_{3}
\end{eqnarray}%
where%
\begin{eqnarray}
c_{21} &=&-\frac{y_{6}}{x_{3}}\frac{Y_{5}}{X_{2}}(c_{11}-1),\quad c_{22}=%
\frac{x_{4}}{x_{3}}-\frac{y_{6}}{x_{3}}\frac{Y_{5}}{X_{2}}c_{12},\quad
c_{23}=-\frac{y_{7}}{x_{3}}-\frac{y_{6}}{x_{3}}\frac{Y_{5}}{X_{2}}c_{13} 
\nonumber \\
c_{24} &=&-\frac{y_{6}}{x_{3}}\frac{X_{5}}{X_{2}}-\frac{y_{6}}{x_{3}}\frac{%
Y_{5}}{X_{2}}c_{14},\quad c_{25}=\frac{y_{6}}{x_{3}}\frac{X_{1}}{X_{2}}-%
\frac{y_{6}}{x_{3}}\frac{Y_{5}}{X_{2}}c_{15}  \nonumber \\
c_{26} &=&F_{1}\frac{x_{4}}{x_{3}}\frac{Y_{6}}{X_{2}}+F_{2}\frac{x_{6}}{x_{3}%
}\frac{X_{3}}{X_{2}}+\frac{y_{6}}{x_{3}}\frac{Y_{7}}{X_{2}}-\frac{y_{6}}{%
x_{3}}\frac{Y_{5}}{X_{2}}c_{16}  \nonumber \\
c_{27} &=&F_{3}\frac{x_{6}}{x_{3}}\frac{X_{3}}{X_{2}}+\frac{x_{4}}{x_{3}}%
\frac{Y_{6}}{X_{2}}-\frac{y_{6}}{x_{3}}\frac{Y_{5}}{X_{2}}c_{17},\quad
c_{28}=\frac{x_{6}}{x_{3}}\frac{X_{3}}{X_{2}}  \nonumber \\
c_{29} &=&-f_{1}\frac{y_{7}}{x_{3}}\frac{Y_{6}}{X_{2}}-f_{1}F_{1}\frac{y_{6}%
}{x_{3}}\frac{X_{4}}{X_{2}}-f_{1}F_{2}\frac{X_{6}}{X_{2}}-\frac{y_{6}}{x_{3}}%
\frac{Y_{5}}{X_{2}}c_{18}  \nonumber \\
c_{210} &=&-\frac{y_{7}}{x_{3}}\frac{Y_{6}}{X_{2}}-F_{1}\frac{y_{6}}{x_{3}}%
\frac{X_{4}}{X_{2}}-F_{2}\frac{X_{6}}{X_{2}}-\frac{y_{6}}{x_{3}}\frac{Y_{5}}{%
X_{2}}c_{19},\quad  \nonumber \\
c_{211} &=&-f_{1}\frac{y_{6}}{x_{3}}\frac{X_{4}}{X_{2}}-f_{1}F_{3}\frac{X_{6}%
}{X_{2}}-\frac{y_{6}}{x_{3}}\frac{Y_{5}}{X_{2}}c_{110}  \nonumber \\
c_{212} &=&-\frac{y_{6}}{x_{3}}\frac{X_{4}}{X_{2}}-F_{3}\frac{X_{6}}{X_{2}}-%
\frac{y_{6}}{x_{3}}\frac{Y_{5}}{X_{2}}c_{111},\quad c_{213}=-f_{1}\frac{X_{6}%
}{X_{2}},\quad c_{214}=-\frac{X_{6}}{X_{2}}
\end{eqnarray}

\item $(FF_{23},ff_{23})$%
\begin{eqnarray}
D_{3}b_{1}
&=&a_{31}b_{1}D_{3}+a_{32}B_{1}d_{1}+a_{33}B_{1}d_{2}+a_{34}B_{3}d_{1}+a_{35}B_{3}d_{2}+a_{36}B_{2}c_{1}
\nonumber \\
&&+a_{37}B_{2}c_{3}+a_{28}b_{2}C_{1}+a_{29}b_{2}C_{3}  \nonumber \\
C_{1}d_{3}
&=&A_{31}d_{3}C_{1}+A_{32}D_{1}c_{1}+A_{33}D_{2}c_{1}+A_{34}D_{1}c_{3}+A_{35}D_{2}c_{3}+A_{36}B_{1}c_{2}
\nonumber \\
&&+A_{37}B_{3}c_{2}+A_{28}b_{1}C_{2}+A_{29}b_{3}C_{2}
\end{eqnarray}%
where%
\begin{eqnarray}
a_{31} &=&\frac{x_{2}}{x_{3}}\frac{X_{2}}{X_{3}}+\frac{x_{2}}{x_{3}}\frac{%
Y_{6}}{X_{3}}X_{15}  \nonumber \\
a_{32} &=&-f_{1}\frac{y_{7}}{x_{3}}\frac{Y_{5}}{X_{3}}+\frac{y_{5}}{x_{3}}%
\frac{Y_{7}}{X_{3}}A_{11}-(Q_{1})a_{22}-(Q_{2})a_{12}+\frac{x_{2}}{x_{3}}%
\frac{Y_{6}}{X_{3}}X_{12}  \nonumber \\
a_{33} &=&-\frac{y_{7}}{x_{3}}\frac{Y_{5}}{X_{3}}-(Q_{2})a_{13}-(Q_{1})a_{28}
\nonumber \\
a_{34} &=&-f_{1}\frac{y_{6}}{x_{3}}\frac{X_{2}}{X_{3}}-(Q_{1})a_{24}+\frac{%
x_{2}}{x_{3}}\frac{Y_{6}}{X_{3}}X_{11},\quad a_{35}=-\frac{y_{6}}{x_{3}}%
\frac{X_{2}}{X_{3}}-(Q_{1})a_{25}  \nonumber \\
a_{36} &=&\frac{y_{5}}{x_{3}}\frac{Y_{7}}{X_{3}}%
A_{16}-(Q_{1})a_{26}-(Q_{2})a_{14}+\frac{x_{2}}{x_{3}}\frac{Y_{6}}{X_{3}}%
X_{18}  \nonumber \\
a_{37} &=&-\frac{y_{7}}{x_{3}}\frac{X_{1}}{X_{3}}-(Q_{2})a_{15}-(Q_{1})a_{27}
\nonumber \\
a_{38} &=&\frac{y_{5}}{x_{3}}\frac{Y_{7}}{X_{3}}%
A_{14}-(Q_{2})a_{16}-(Q_{1})a_{28}+\frac{x_{2}}{x_{3}}\frac{Y_{6}}{X_{3}}%
X_{16}  \nonumber \\
a_{39} &=&\frac{y_{5}}{x_{3}}\frac{X_{1}}{X_{3}}+\frac{y_{5}}{x_{3}}\frac{%
Y_{7}}{X_{3}}A_{15}-(Q_{1})a_{29}+\frac{x_{2}}{x_{3}}\frac{Y_{6}}{X_{3}}%
X_{17}
\end{eqnarray}

\item $(FF_{13},ff_{13})$%
\begin{eqnarray}
D_{3}b_{2}
&=&b_{31}B_{2}d_{3}+b_{32}B_{2}d_{1}+b_{33}B_{2}d_{2}+b_{34}b_{2}D_{3}+b_{35}B_{1}b_{1}+b_{36}B_{1}b_{3}
\nonumber \\
&&+b_{37}B_{3}b_{1}+b_{38}B_{3}b_{3}  \nonumber \\
C_{2}d_{3}
&=&B_{31}D_{3}c_{2}+B_{32}D_{1}c_{2}+B_{33}D_{2}c_{2}+B_{34}d_{3}C_{2}+B_{35}C_{1}c_{1}+B_{36}C_{3}c_{1}
\nonumber \\
&&+B_{37}C_{1}c_{3}+B_{38}C_{3}c_{3}
\end{eqnarray}%
where%
\begin{eqnarray}
b_{31} &=&-\frac{y_{7}}{x_{3}}\frac{X_{1}}{X_{3}}-(Q_{2})b_{14}-(Q_{1})b_{24}
\nonumber \\
b_{32} &=&-f_{2}\frac{y_{7}}{x_{3}}\frac{X_{1}}{X_{3}}+\frac{x_{1}}{x_{3}}%
\frac{Y_{5}}{X_{3}}D_{13}+\frac{x_{1}}{x_{3}}\frac{Y_{7}}{X_{3}}%
(B_{11}+D_{13}B_{16}+E_{03}B_{15})  \nonumber \\
&&-(Q_{2})b_{12}-(Q_{1})b_{22}  \nonumber \\
b_{33} &=&-f_{3}\frac{y_{7}}{x_{3}}\frac{X_{1}}{X_{3}}+\frac{x_{1}}{x_{3}}%
\frac{Y_{5}}{X_{3}}D_{14}+\frac{x_{1}}{x_{3}}\frac{Y_{7}}{X_{3}}%
(D_{14}B_{16}+E_{02}B_{15})-(Q_{2})b_{13}-(Q_{1})b_{23}  \nonumber \\
b_{34} &=&\frac{x_{1}}{x_{3}}\frac{X_{1}}{X_{3}}+\frac{x_{1}}{x_{3}}\frac{%
Y_{7}}{X_{3}}(B_{14}+D_{17}B_{16})+\frac{x_{1}}{x_{3}}\frac{Y_{5}}{X_{3}}%
D_{17}  \nonumber \\
b_{35} &=&\frac{x_{1}}{x_{3}}\frac{Y_{5}}{X_{3}}D_{12}+\frac{x_{1}}{x_{3}}%
\frac{Y_{7}}{X_{3}}(D_{12}B_{16}+E_{01}B_{15})-(Q_{2})b_{15}-(Q_{1})b_{25} 
\nonumber \\
b_{36} &=&-\frac{y_{7}}{x_{3}}\frac{Y_{5}}{X_{3}}-(Q_{2})b_{16}-(Q_{1})b_{26}
\nonumber \\
b_{37} &=&\frac{x_{1}}{x_{3}}\frac{Y_{5}}{X_{3}}D_{11}+\frac{x_{1}}{x_{3}}%
\frac{Y_{7}}{X_{3}}D_{11}B_{16}-(Q_{1})b_{27}  \nonumber \\
b_{28} &=&-\frac{y_{6}}{x_{3}}\frac{X_{2}}{X_{3}}-(Q_{1})b_{28}
\end{eqnarray}%
Here we have used the notation%
\begin{equation}
Q_{1}=F_{3}+\frac{y_{6}}{x_{3}}\frac{Y_{6}}{X_{3}},\qquad Q_{2}=F_{2}+F_{1}%
\frac{y_{6}}{x_{3}}\frac{Y_{6}}{X_{3}}+\frac{y_{7}}{x_{3}}\frac{Y_{7}}{X_{3}}
\end{equation}%
The remaining commutation relations do not participate effectively for the
algebraic {\small BA}. Nevertheless, we will write them below, but without
their coefficients which are very cumbersome.

\item $(FF_{25},ff_{25})$ and $(F_{27},f_{27})$ 
\begin{eqnarray}
D_{2}b_{3}
&=&X_{21}b_{3}D_{2}+X_{22}b_{1}D_{2}+X_{23}B_{1}d_{1}+X_{24}B_{1}d_{2}+X_{25}B_{1}d_{3}
\nonumber \\
&&+X_{26}B_{3}d_{1}+X_{27}B_{3}d_{2}+X_{28}B_{3}d_{3}+X_{29}B_{2}c_{1}+X_{210}B_{2}c_{3}
\nonumber \\
&&+X_{211}b_{2}C_{1}+X_{212}b_{2}C_{3}  \nonumber \\
C_{3}d_{2}
&=&x_{21}d_{2}C_{3}+x_{21}d_{2}C_{1}+x_{23}D_{1}c_{1}+x_{24}D_{2}c_{1}+x_{25}D_{3}c_{1}
\nonumber \\
&&+x_{26}D_{1}c_{3}+x_{27}D_{2}c_{3}+x_{28}D_{3}c_{3}+x_{29}B_{1}c_{2}+x_{210}B_{3}c_{2}
\nonumber \\
&&+x_{211}b_{1}C_{2}+x_{212}b_{3}C_{2}
\end{eqnarray}%
and%
\begin{eqnarray}
C_{1}b_{2}
&=&Y_{11}b_{2}C_{1}+Y_{12}b_{2}C_{3}+Y_{13}B_{2}c_{1}+Y_{14}B_{2}c_{3}+Y_{15}b_{1}D_{1}
\nonumber \\
&&+Y_{16}b_{1}D_{2}+Y_{17}b_{3}D_{1}+Y_{18}b_{3}D_{2}+Y_{19}B_{1}d_{1}+Y_{110}B_{1}d_{2}
\nonumber \\
&&+Y_{111}B_{1}d_{3}+Y_{112}B_{3}d_{1}+Y_{113}B_{3}d_{2}+Y_{114}B_{3}d_{3} 
\nonumber \\
C_{2}b_{1}
&=&y_{11}b_{1}C_{2}+y_{12}b_{3}C_{2}+y_{13}B_{1}c_{2}+y_{14}B_{3}c_{2}+y_{15}d_{1}C_{1}
\nonumber \\
&&+y_{16}d_{2}C_{1}+y_{17}d_{1}C_{3}+y_{18}d_{2}C_{3}+y_{19}D_{1}c_{1}+y_{110}D_{2}c_{1}
\nonumber \\
&&+y_{111}D_{3}c_{1}+y_{112}D_{1}c_{3}+y_{113}D_{2}c_{3}+y_{114}D_{3}c_{3}
\end{eqnarray}

\item $(FF_{24},ff_{24})$ and $(F_{37},f_{37})$ 
\begin{eqnarray}
C_{2}b_{2}
&=&c_{31}D_{2}d_{3}+c_{32}B_{2}c_{2}+c_{33}D_{3}d_{3}+c_{34}D_{2}d_{2}+c_{35}B_{3}c_{1}
\nonumber \\
&&+c_{36}d_{2}D_{3}+c_{37}D_{1}d_{2}+c_{38}d_{1}D_{3}+c_{39}d_{1}D_{2}+c_{310}b_{3}C_{1}
\nonumber \\
&&+c_{311}B_{1}c_{1}+c_{312}D_{3}d_{1}+c_{313}d_{3}D_{1}+c_{314}d_{2}D_{2}+c_{315}D_{3}d_{2}
\nonumber \\
&&+c_{316}D_{1}d_{3}+c_{317}b_{2}C_{2}+c_{318}d_{1}D_{1}+c_{319}D_{1}d_{1}+c_{320}D_{2}d_{1}
\nonumber \\
&&+c_{321}b_{3}C_{3}+c_{322}b_{1}C_{1}+c_{323}d_{2}D_{1}+c_{324}b_{1}C_{3}+c_{325}B_{3}c_{3}
\end{eqnarray}%
and%
\begin{eqnarray}
C_{3}b_{3}
&=&c_{41}d_{1}D_{2}+c_{42}D_{1}d_{2}+c_{43}D_{2}d_{1}+c_{44}B_{1}c_{1}+c_{45}d_{2}D_{3}
\nonumber \\
&&+c_{46}b_{3}C_{1}+c_{47}b_{2}C_{2}+c_{48}D_{3}d_{1}+c_{49}d_{2}D_{1}+c_{410}D_{3}d_{2}
\nonumber \\
&&+c_{411}b_{1}C_{3}+c_{412}d_{2}D_{2}+c_{413}b_{1}C_{1}+c_{414}D_{1}d_{3}+c_{415}d_{1}D_{3}
\nonumber \\
&&+c_{416}D_{2}d_{2}+c_{417}d_{1}D_{1}+c_{418}D_{1}d_{1}+c_{419}B_{2}c_{2}+c_{420}d_{3}D_{1}
\nonumber \\
&&+c_{421}B_{3}c_{3}+c_{422}b_{3}C_{3}+c_{423}B_{3}c_{1}+c_{424}D_{3}d_{3}+c_{425}D_{2}d_{3}
\end{eqnarray}

\item $(FF_{12},ff_{12})$ and $(FF_{45},ff_{45})$ 
\begin{eqnarray}
D_{3}b_{3}
&=&X_{31}b_{3}D_{3}+X_{32}b_{1}D_{3}+X_{33}B_{1}d_{1}+X_{34}B_{1}d_{2}+X_{35}B_{1}d_{3}
\nonumber \\
&&+X_{36}B_{3}d_{1}+X_{37}B_{3}d_{2}+X_{38}B_{3}d_{3}+X_{39}B_{2}c_{1}+X_{310}B_{2}c_{3}
\nonumber \\
&&+X_{311}b_{2}C_{1}+X_{312}b_{2}C_{3}  \nonumber \\
C_{3}d_{3}
&=&x_{31}d_{3}C_{3}+x_{32}d_{3}C_{1}+x_{33}D_{1}c_{1}+x_{34}D_{2}c_{1}+x_{35}D_{3}c_{1}
\nonumber \\
&&+x_{36}D_{1}c_{3}+x_{37}D_{2}c_{3}+x_{38}D_{3}c_{3}+x_{39}B_{1}c_{2}+x_{310}B_{3}c_{2}
\nonumber \\
&&+x_{311}b_{1}C_{2}+x_{312}b_{3}C_{2}
\end{eqnarray}%
and%
\begin{eqnarray}
C_{3}b_{2}
&=&Y_{21}b_{2}C_{3}+Y_{22}b_{2}C_{1}+Y_{23}B_{2}c_{1}+Y_{24}B_{2}c_{3}+Y_{25}b_{1}D_{1}+Y_{26}b_{1}D_{2}
\nonumber \\
&&+Y_{27}b_{1}D_{3}+Y_{28}b_{3}D_{1}+Y_{29}b_{3}D_{2}+Y_{210}b_{3}D_{3}+Y_{211}B_{1}d_{1}+Y_{212}B_{1}d_{2}
\nonumber \\
&&+Y_{213}B_{1}d_{3}+Y_{214}B_{3}d_{1}+Y_{215}B_{3}d_{2}+Y_{216}B_{3}d_{3} 
\nonumber \\
C_{2}b_{3}
&=&y_{21}b_{3}C_{2}+y_{22}b_{1}C_{2}+y_{23}B_{1}c_{2}+y_{24}B_{3}c_{2}+y_{25}d_{1}C_{1}+y_{26}d_{2}C_{1}
\nonumber \\
&&+y_{27}d_{3}C_{1}+y_{28}d_{1}C_{3}+y_{29}d_{2}C_{3}+y_{210}d_{3}C_{3}+y_{211}D_{1}c_{1}+y_{212}D_{2}c_{1}
\nonumber \\
&&+y_{213}D_{3}c_{1}+y_{214}D_{1}c_{3}+y_{215}D_{2}c_{3}+y_{216}D_{3}c_{3}
\end{eqnarray}
\end{itemize}

\bigskip

\bigskip

\bigskip

\bigskip

\bigskip \bigskip

\bigskip

\bigskip

\bigskip

\bigskip

\end{document}